\documentclass[aps,prl,floatfix,twocolumn,twoside,superscriptaddress,10pt]{revtex4-1}

\usepackage[utf8]{inputenc}
\usepackage[english]{babel}
\usepackage{amsmath,amssymb,graphicx,enumerate,bm}
\usepackage{graphicx} 
\usepackage{subfigure} 
\usepackage{color}
\usepackage{multirow}

\newcommand{\tw}{{t_\mathrm{w}}}
\newcommand{\NR}{{N_\mathrm{R}}}
\newcommand{\bsym}[1]{{\boldsymbol #1}}

\begin{document}

\title{Multifractality in spin glasses}


\author{M.~Baity-Jesi}\affiliation{Eawag, Überlandstrasse 133, CH-8600 Dübendorf, Switzerland}

\author{E.~Calore}\affiliation{Dipartimento di Fisica e Scienze della Terra,
  Universit\`a di Ferrara and INFN, Via Giuseppe Saragat, 1, 44122 Ferrara,
  Italy}

\author{A.~Cruz}\affiliation{Departamento de F\'{\i}sica Te\'orica,
  Universidad de Zaragoza, 50009 Zaragoza, Spain}\affiliation{Instituto de
  Biocomputac\'{\i}on y F\'{\i}sica de Sistemas Complejos (BIFI), 50018
  Zaragoza, Spain}

\author{L.A.~Fernandez}\affiliation{Departamento de F\'{\i}sica Te\'orica,
  Universidad Complutense, 28040 Madrid, Spain}\affiliation{Instituto de
  Biocomputac\'{\i}on y F\'{\i}sica de Sistemas Complejos (BIFI), 50018
  Zaragoza, Spain}

\author{J.M.~Gil-Narvion}\affiliation{Instituto de Biocomputac\'{\i}on y
  F\'{\i}sica de Sistemas Complejos (BIFI), 50018 Zaragoza, Spain}

\author{I.~Gonzalez-Adalid Pemartin}\affiliation{Departamento de F\'{\i}sica
  Te\'orica, Universidad Complutense, 28040 Madrid, Spain}\email{isidorog@ucm.es}

\author{A.~Gordillo-Guerrero}\affiliation{Departamento de Ingenier\'{\i}a
  El\'ectrica, Electr\'onica y Autom\'atica, U. de Extremadura, 10003,
  C\'aceres, Spain}\affiliation{Instituto de Computac\'{\i}on Cient\'{\i}fica
  Avanzada (ICCAEx), Universidad de Extremadura, 06006 Badajoz,
  Spain}\affiliation{Instituto de Biocomputac\'{\i}on y F\'{\i}sica de Sistemas
  Complejos (BIFI), 50018 Zaragoza, Spain}

\author{D.~I\~niguez}\affiliation{Instituto de Biocomputac\'{\i}on y
  F\'{\i}sica de Sistemas Complejos (BIFI), 50018 Zaragoza,
  Spain}\affiliation{Fundaci\'on ARAID, Diputac\'{\i}on General de Arag\'on,
  50018 Zaragoza, Spain}\affiliation{Departamento de F\'{\i}sica Te\'orica,
  Universidad de Zaragoza, 50009 Zaragoza, Spain}

\author{A.~Maiorano}\affiliation{Dipartimento di
  Biotecnologie, Chimica e Farmacia, Universit\`a degli studi di Siena, 3100
  Siena, Italy and INFN, Sezione di Roma 1, 00185 Rome,
  Italy}\affiliation{Instituto de Biocomputac\'{\i}on y F\'{\i}sica de Sistemas
  Complejos (BIFI), 50018 Zaragoza, Spain}

\author{E.~Marinari}\affiliation{Dipartimento di Fisica, Sapienza Universit\`a
  di Roma, and CNR-Nanotec, Rome unit and INFN, Sezione di Roma 1, 00185 Rome,
  Italy}

\author{V.~Martin-Mayor}\affiliation{Departamento de F\'{\i}sica
  Te\'orica, Universidad Complutense, 28040 Madrid, Spain}\affiliation{Instituto
  de Biocomputac\'{\i}on y F\'{\i}sica de Sistemas Complejos (BIFI), 50018
  Zaragoza, Spain}

\author{J.~Moreno-Gordo}\affiliation{Instituto de Biocomputac\'{\i}on y
  F\'{\i}sica de Sistemas Complejos (BIFI), 50018 Zaragoza,
  Spain}\affiliation{Departamento de F\'{\i}sica Te\'orica, Universidad de
  Zaragoza, 50009 Zaragoza, Spain}\affiliation{Departamento de F\'{\i}sica,
  Universidad de Extremadura, 06006 Badajoz, Spain}\affiliation{Instituto de
  Computac\'{\i}on Cient\'{\i}fica Avanzada (ICCAEx), Universidad de
  Extremadura, 06006 Badajoz, Spain}

\author{A.~Mu\~noz~Sudupe}\affiliation{Departamento de F\'{\i}sica
  Te\'orica, Universidad Complutense, 28040 Madrid, Spain}\affiliation{Instituto
  de Biocomputac\'{\i}on y F\'{\i}sica de Sistemas Complejos (BIFI), 50018
  Zaragoza, Spain}

\author{D.~Navarro}\affiliation{Fundaci\'on ARAID, Diputac\'{\i}on General de
  Arag\'on, 50018 Zaragoza, Spain}

\author{I.~Paga}\affiliation{Institute of Nanotechnology, Consiglio Nazionale
  delle Ricerche (CNR-NANOTEC), Piazzale Aldo Moro 5, I-00185 Rome, Italy}

\author{G.~Parisi}\affiliation{Dipartimento di Fisica, Sapienza Universit\`a
  di Roma, and CNR-Nanotec, Rome unit and INFN, Sezione di Roma 1, 00185 Rome,
  Italy}

\author{S.~Perez-Gaviro}\affiliation{Departamento de F\'{\i}sica Te\'orica,
  Universidad de Zaragoza, 50009 Zaragoza, Spain}\affiliation{Instituto de
  Biocomputac\'{\i}on y F\'{\i}sica de Sistemas Complejos (BIFI), 50018
  Zaragoza, Spain}

\author{F.~Ricci-Tersenghi}\affiliation{Dipartimento di Fisica, Sapienza
  Universit\`a di Roma, and CNR-Nanotec, Rome unit and INFN, Sezione di Roma
  1, 00185 Rome, Italy}

\author{J.J.~Ruiz-Lorenzo}\affiliation{Departamento de F\'{\i}sica,
  Universidad de Extremadura, 06006 Badajoz, Spain}\affiliation{Instituto de
  Computac\'{\i}on Cient\'{\i}fica Avanzada (ICCAEx), Universidad de
  Extremadura, 06006 Badajoz, Spain}\affiliation{Instituto de Biocomputac\'{\i}on
  y F\'{\i}sica de Sistemas Complejos (BIFI), 50018 Zaragoza, Spain}

\author{S.F.~Schifano}\affiliation{Dipartimento di Scienze dell'Ambiente e
  della Prevenzione Università di Ferrara e INFN Sezione di Ferrara, I-44122
  Ferrara, Italy}

\author{B.~Seoane}\affiliation{Universit\'e Paris-Saclay, CNRS, INRIA Tau
  team, LISN, 91190, Gif-sur-Yvette, France}\affiliation{Instituto de
  Biocomputac\'{\i}on y F\'{\i}sica de Sistemas Complejos (BIFI), 50018
  Zaragoza, Spain}

\author{A.~Tarancon}\affiliation{Departamento de F\'{\i}sica Te\'orica,
  Universidad de Zaragoza, 50009 Zaragoza, Spain}\affiliation{Instituto de
  Biocomputac\'{\i}on y F\'{\i}sica de Sistemas Complejos (BIFI), 50018
  Zaragoza, Spain}

\author{D.~Yllanes}\affiliation{Chan Zuckerberg Biohub --- San Francisco,
  San Francisco, CA, 94158}\affiliation{Instituto de Biocomputac\'{\i}on y
  F\'{\i}sica de Sistemas Complejos (BIFI), 50018 Zaragoza, Spain}

\collaboration{Janus Collaboration}
\noaffiliation
\date{\today}



\begin{abstract}
We unveil the multifractal behavior of Ising spin glasses in their
low-temperature phase. Using the Janus~II custom-built supercomputer, the
spin-glass correlation function is studied locally. Dramatic fluctuations are
found when pairs of sites at the same distance are compared. The scaling of
these fluctuations, as the spin-glass coherence length grows with time, is
characterized through the computation of the singularity spectrum and its
corresponding Legendre transform. A comparatively small number of site pairs
controls the average correlation that governs the response to a magnetic
field. We explain how this scenario of dramatic fluctuations (at length scales
smaller than the coherence length) can be reconciled with the smooth,
self-averaging behavior that has long been considered to describe spin-glass
dynamics.
\end{abstract}

\maketitle

The notion of multifractality~\cite{frisch:85,harte:01} refers to
situations where many different fractal behaviors coexist within the same
system. A major role is played in this context by scale symmetry, see,
\emph{e.g.},~\cite{wilson:79,parisi:88,barnsley:12}: in many situations in
physics, chemistry and beyond, apparently random objects look the same when
the observation scale is changed. The scale change is often quantitatively
characterized through a number, the fractal dimension. Multifractals (as
opposed to fractals) are systems that need many fractal dimensions to get
their scaling properties fully characterized.

Some of the first examples of multifractal behavior appeared in physics, in
the contexts of turbulence~\cite{benzi:84}, Anderson
localization~\cite{castellani:86} and diffusion-limited
aggregates~\cite{stanley:88}. A unifying language was soon introduced in a
study of chaotic dynamics~\cite{halsey:86,halsey:86b}. The concept has gained
popularity as the list of systems exhibiting some form of multifractality has
steadily grown. To name only a few, let us recall surface
growth~\cite{barabasi:09}, human heartbeat dynamics~\cite{ivanov:99}, mating
copepods~\cite{seuront:14,klopper:14}, rainfall~\cite{deidda:00} or the
analysis of financial time series~\cite{alvarez:08}.

Here we add a (perhaps) surprising member to the list: the off-equilibrium
dynamics of spin-glass systems~\cite{mydosh:93,charbonneau:23}. These
disordered magnetic alloys have long been regarded as a paradigmatic toy model
for the study of glassiness, optimization, biology, financial markets or
social dynamics. It is surprising that such a prominent feature as
multifractality has gone unnoticed for such a well-studied model.

\begin{figure}
\centering
\includegraphics[width=\linewidth]{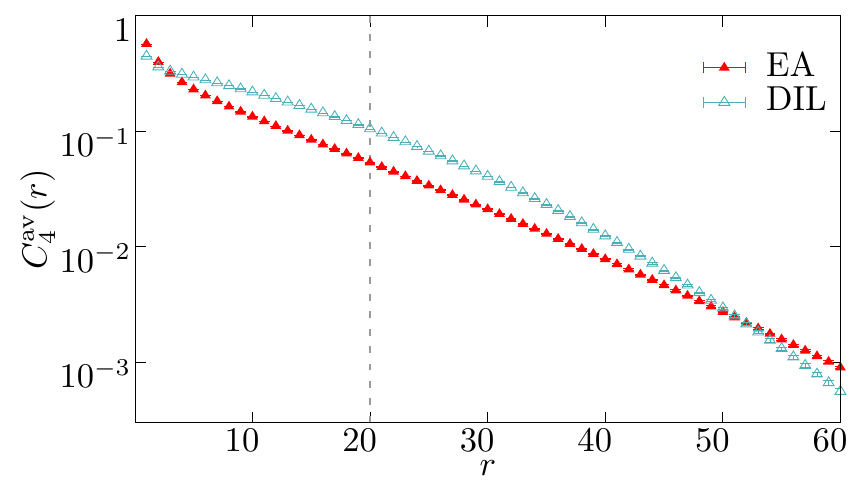}
\caption{The correlation function, \eqref{eq:C4-av}, as computed for the
  three-dimensional Ising diluted ferromagnet (DIL) and for the Ising spin
  glass (EA), versus distance $r$. Data were obtained in systems of linear
  size $L=160$ with coherence length $\xi(\tw)=20$ (dashed vertical line) at
  temperature $T=0.9$---recall that $T_{\mathrm{c}}\approx 1.1$ for
  EA~\cite{janus:13}. As explained in \textbf{Methods}, the coherence length
  is computed from the integral $I_2=\int_0^\infty
  r^2C^{\text{av}}_4(r)\mathrm{d}r$ (the integrand is shown in the Supporting
  Information).  }
\label{fig:C4_EA_vs_ISDIL}
\end{figure}

The explanation for the above paradox rests on the finite coherence length
$\xi(\tw)$ that develops when a spin glass, initially at some very high
temperature, is suddenly cooled below the critical temperature $T_\mathrm{c}$,
and let to relax for a waiting time $\tw$---most experimental work on spin
glasses is carried out under non-equilibrium conditions~\cite{vincent:97}. As
$\tw$ increases, glassy domains of growing size $\xi(\tw)$ develop, see
Fig.~\ref{fig:C4_EA_vs_ISDIL}. The growth of $\xi(\tw)$ is sluggish for a spin
glass, reaching only $\xi\sim$~200 lattice spacings for $\tw
\sim$1~hour~\cite{zhai:19,zhai-janus:20a}. Now, when one measures the magnetic
response to an external field, which is the main experimental probe of
spin-glass dynamics, an average over the whole sample is carried out. Since
the sample is effectively composed of many independent domains of linear size
$\sim\xi(\tw)$, the central limit theorem eliminates from the average response
the large fluctuations that could ultimately cause multifractal behavior. With
few exceptions (see below), most numerical work has emphasized the
space-averaged correlation function in Fig.~\ref{fig:C4_EA_vs_ISDIL}. Besides,
see \textbf{Methods}, studying correlations without spatial averages is very
demanding computationally.

It follows from the above considerations that multifractal behavior in spin
glasses should be investigated in large statistical deviations that occur at a
length scale smaller than (or comparable to) $\xi(\tw)$, definitively not the
standard framework either for experiments\citep[see][for
  instance]{zhai-janus:20a,zhai-janus:21,zhai:22} or for
simulations~\cite{janus:08,janus:17,janus:18}. There is, however, an important
exception. Recently, progress has been achieved~\cite{janus:23} in the
theoretical interpretation of the experimental rejuvenation and memory effects
in spin glasses~\cite{jonason:98}. Crucial for this achievement was the study
of temperature chaos in the off-equilibrium dynamics \emph{at the $\xi(\tw)$
  length scale}~\cite{janus:21}, through numerical simulations using the Janus
II dedicated supercomputer~\cite{janus:14}. As we shall show below, the
consideration of fluctuations at the $\xi(\tw)$ length scale still holds
surprises.

Specifically, we shall consider the spin-glass correlation function, see
\textbf{Methods} for definitions. The space-averaged correlation function is a
well-known quantity and the basis for the computation of $\xi(\tw)$ explained
in Fig.~\ref{fig:C4_EA_vs_ISDIL}.  We shall depart from the standard approach,
however, by avoiding the spatial average. We shall compute the correlation
function for a pair of sites ${\bsym{x}}$ and ${\bsym{y}}$, and consider the
statistical fluctuations induced by varying ${\bsym{x}}$ while fixing
$\bsym{r}=\bsym{y}-\bsym{x}$.

The reader may argue that it is difficult to find large statistical
fluctuations in a mathematical object bounded between 0 and 1, such as the
spin-glass correlation function.  A moment of thought will reveal that large
fluctuations are only possible if the average of the correlation function goes
to zero as $\xi(\tw)$ grows, so that the correlation function at a given site
can get large if measured in units of the averaged correlation.

\begin{figure}[t]
\centering
\includegraphics[width=\linewidth]{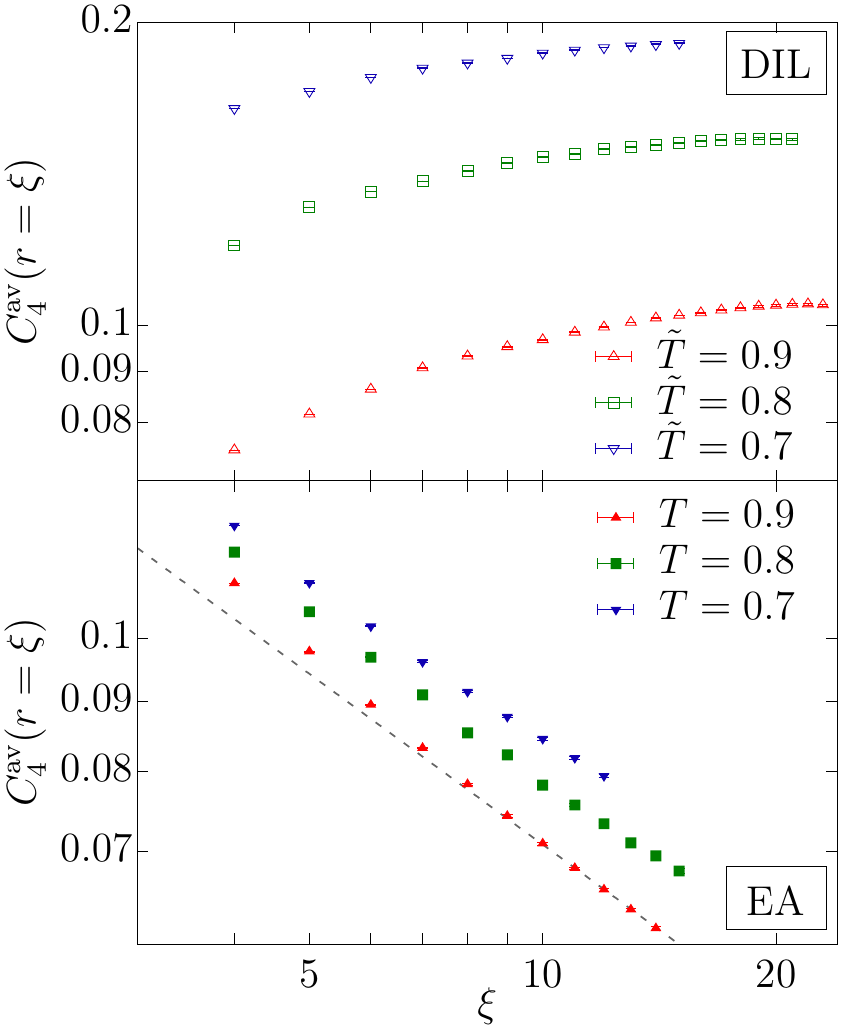}
\caption{Correlation function $C_4^{\text{av}}(r=\xi(\tw))$, see
  ~\eqref{eq:C4-av} versus the coherence length $\xi(\tw)$, as computed for
  DIL (\textbf{top}) and for EA (\textbf{bottom}) at temperatures
  $T,\ \tilde{T} = 0.9,0.8$ and $0.7$ (see \textbf{Methods} for a complete
  definition of $\tilde{T}$). Error bars are smaller than the point size. The
  dashed line is our fit to~\eqref{eq:tau-q-def}, with $q=1$, for EA at
  $T=0.9$ (to avoid scaling corrections, we fit in the range $\xi(\tw)\in
  [10,20]$, see SI for further information). Note that, while the DIL
  $C_4^{\text{av}}(r=\xi(\tw))$ tends to a $T$-dependent positive limit for
  large coherence length (which excludes multiscaling at $T<T_{\text{c}}$),
  the spin-glass correlation functions steadily decrease with $\xi(\tw)$.}
\label{fig:C4_vs_xi}
\end{figure}

Indeed, spin glasses are peculiar among systems with domain-growth
off-equilibrium dynamics. Fig.~\ref{fig:C4_vs_xi} compares the space-averaged
correlation function at distance $r=\xi(\tw)$ for two Ising systems in space
dimension $D=3$, the link-diluted ferromagnet and the spin glass. In the
ferromagnet, the correlation function goes to a constant value (the squared
spontaneous magnetization) as $\xi$ grows. Hence, large deviations and
multifractality are possible for the ferromagnet only at $T_\mathrm{c}$, where
the spontaneous magnetization vanishes~\footnote{At its critical temperature, the
  two-dimensional diluted ferromagnetic Potts model with more than two
  states presents multiscaling as well~\cite{ludwig:90} ---this is
  also the case for the diluted Ising model in
  $D=3$~\cite{davis:00,marinari:23}.}.  In the spin glass, instead, the
correlation function scales as $1/r^\theta$ for distances up to $r\sim
\xi$~\footnote{The droplet picture of spin
  glasses~\cite{mcmillan:83,bray:78,fisher:86} predicts $\theta\!=\!0$,
  similarly to the ferromagnet. Neither simulations nor experimental data are
  compatible with $\theta\!=\!0$, unless one is willing to accept that the
  available range of $\xi(\tw)$ is too small to display the true asymptotic
  behavior~\cite{janus:18}.}. Hence, unlike the diluted ferromagnet, the spin
glass can accommodate large fluctuations for all $T<T_{\mathrm{c}}$. This is
why here we decide to focus on the spin glass.

\section*{Results}

\begin{figure}
\centering
\includegraphics[width=\linewidth]{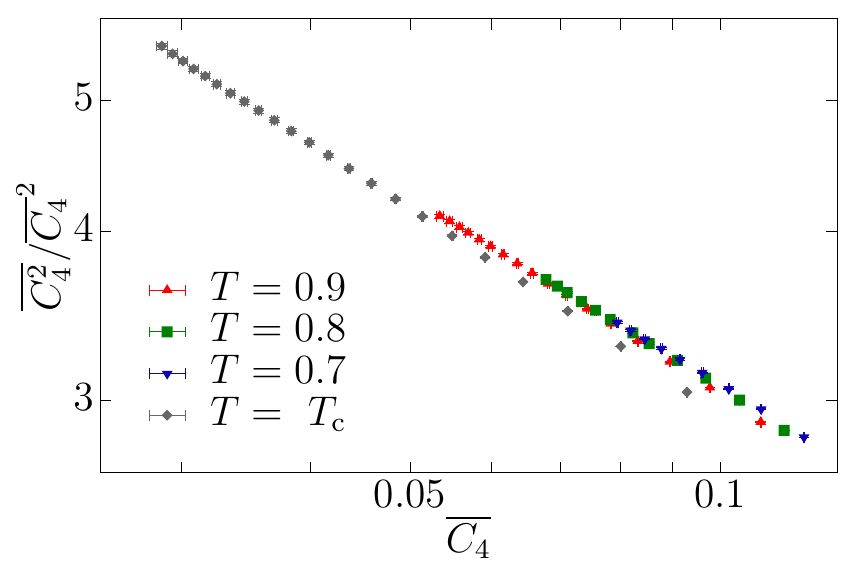}
\caption{Ratio of the second moment of the spin-glass correlation function
  $C_4$ computed at $r=\xi(\tw)$, $\overline{C_4^2}$, to the squared first
  moment, $\overline{C_4}^2$, as a function of $\overline{C_4}$. We show the
  data for all temperatures considered in this work. $\overline{C_4}$ tends to
  zero as the coherence length $\xi(\tw)$ gets large, recall
  Fig.~\ref{fig:C4_vs_xi}. Note that $\overline{C_4^2}/\overline{C_4}^2$
  scales with $\overline{C_4}$ as a power law, which indicates that in the
  scaling limit (\emph{i.e.}, $\xi(\tw)\to\infty$ or $\overline{C_4}\to 0$)
  the order of magnitude of $\overline{C_4^2}$ is larger than the one of
  $\overline{C_4}^2$. Data in the glassy phase, $T<T_\mathrm{c}$, roughly
  follow the same scaling curve. At the critical point there is still a power
  type relation with a slightly different exponent. Error bars are smaller
  than the points size. The same data are shown as a function of $\xi(\tw)$ in
  the SI.}
\label{fig:C4_norm_vs_C4}
\end{figure}

The first indication of large deviations in the statistics of the spin-glass
correlation function $C_4$ is shown in Fig.~\ref{fig:C4_norm_vs_C4}, where we
select the distance $r=\xi(\tw)$. The ratio of the second moment of $C_4$,
$\overline{C_4^2}$, to the first moment squared, $\overline{C_4}^2$, nicely
follows a power law as a function of $\overline{C_4}$ \citep[this type of
  analysis was pioneered in][]{benzi:93}. If continued to $\overline{C_4}\to
0$ (\emph{i.e.}, as $\xi(\tw)$ grows, see Fig.~\ref{fig:C4_vs_xi}), this power
law implies that the orders of magnitude of $\overline{C_4^2}$ and
$\overline{C_4}^2$ differ in the large-$\xi(\tw)$ scaling limit. This behavior
is not reminiscent of a monofractal, which in the scaling limit is
characterized by a single quantity (say, $\overline{C_4}$).
\begin{figure*}[bt!]
\centering
\includegraphics[width=\linewidth]{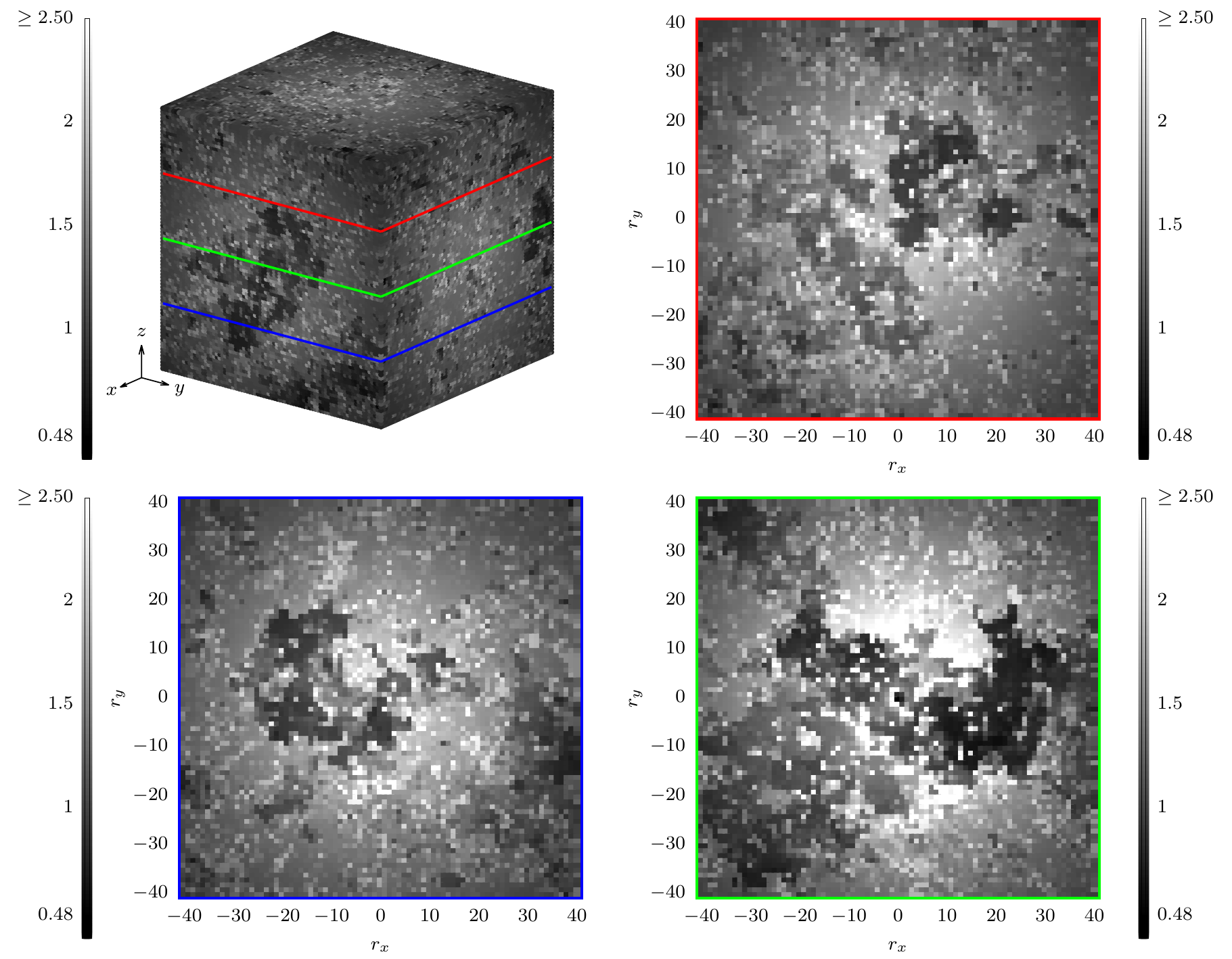}
\caption{Grayscale representation of the order-of-magnitude modulating factor
  $M(\bsym{x},\bsym{r},\tw)$, see~\eqref{eq:M-def}, computed for site
  $\bsym{x}=(64,64,64)$ of a sample with coherence length $\xi(\tw)=20$, at
  $T=0.9$, with an $\NR=512$ estimator (see \textbf{Methods}). We show results
  for displacement vectors $\bsym{r}=(r_x,r_y,r_z)$ in a cube $-40\leq
  r_x,r_y,r_z\leq 40$. The top-left panel depicts the three visible faces of
  the cube, while the other three panels show sections at $r_z=-20,0,20$,
  respectively. Our color code is darker the smaller
  $M(\bsym{x},\bsym{r},\tw)$ (hence, the more slowly correlations decay with
  distance).  For ease of representation, we have chosen a color code linear
  between the minimal value of $M(\bsym{x},\bsym{r},\tw)$ and 2.5.
  Displacements $\bsym{r}$ with $M(\bsym{x},\bsym{r},\tw)>2.5$ are depicted as
  if $M(\bsym{x},\bsym{r},\tw)=2.5$. See SI for more examples of this
  modulating factor.}
\label{fig:cube1}
\end{figure*}

We also note from Fig.~\ref{fig:C4_norm_vs_C4} that all our data with
$T<T_\mathrm{c}$ follow the same scaling curve, which slightly differs
from its counterpart at the critical point. This is not completely
unexpected, because the $\epsilon$-expansion tells us that the average
\protect{$C_4$} at $T_{\text{c}}$ decays as a power law with distance
with an exponent~\cite{chen:77} that is twice as large as the exponent
for $T<T_{\text{c}}$~\cite{dedominicis:89}. In fact, we lack an
explanation for the similarity of the two exponents that can be
observed in Fig.~\ref{fig:C4_norm_vs_C4}. From now on, our analysis
will focus on our data at $T=0.9$, namely the temperature in the
spin-glass phase where we are able to reach the largest $\xi(\tw)$.

A picture of the physical situation is presented in Fig.~\ref{fig:cube1}. We
may expect a different behavior for the average and the local correlation
function when distances up to $r\sim\xi(\tw)$ are considered
($\theta(T=0.9)\approx 0.4$~\cite{janus:18})~\footnote{The correlation
  function behaves as $C_4^\text{av}(r,\tw)\sim G(r/\xi(\tw))/r^\theta$ for
  large $r$, where the cut-off function $G(x)$ decays faster than
  exponentially as $x$ grows [see, \emph{e.g.},
    Refs.~\cite{janus:08b,fernandez:19}]. Hence, for $r\sim\xi(\tw)$ one may
  consider either power-law scaling in $r$ ---as in~\eqref{eq:M-def}--- or in
  $\xi(\tw)$ ---as in~\eqref{eq:tau-q-def}. The analysis of scale invariance
  in a fractal (or multifractal) geometry typically involves power laws.}:
\begin{equation}\label{eq:M-def}
    C_4^\text{av}(r,\tw)\sim\frac{1}{r^\theta}\,,\quad C_4(\bsym{x},\bsym{x}+\bsym{r};\tw)\sim \frac{1}{r^{\theta M(\bsym{x},\bsym{r},\tw)}}\,.
\end{equation}
As the reader can check from Fig.~\ref{fig:cube1}, the order-of-magnitude
modulating factor $M(\bsym{x},\bsym{r},\tw)$ varies by a factor of 16, which
indicates that there are site pairs $(\bsym{x},\bsym{x}+\bsym{r})$ a lot more
---or a lot less--- correlated than the average. In fact, see
Fig.~\ref{fig:median_vs_C4}, the median correlation function at distance
$r=\xi(\tw)$, scales as $[C_4^\text{av}]^a$, with $a\approx 1.5$. In other
words, the \emph{typical} correlation function is a lot smaller than the
average value.

In order to make the above qualitative description quantitative, we consider
the moments of the probability distribution of $C_4$ at distance
$r=\xi(\tw)$. The $q$-th moment turns out to follow a scaling law
\begin{equation}\label{eq:tau-q-def}
    \overline{C_4^q}\sim \frac{A_q}{\xi^{\tau(q)}}\,.
\end{equation}
Fig.~\ref{fig:tau_q} shows the $\tau(q)$ function, which significantly differs
from the monofractal behavior $\tau^\text{mono}(q)=q\,\tau^\text{mono}(1)$. It
is this departure from linear behavior that justifies using the term
\emph{multifractal} to describe spin-glass dynamics \citep[see,
  \emph{e.g.},][]{halsey:86}.

For large moments, $\tau(q)$ seems to grow as $\log q$ (see the inset in
Fig.~\ref{fig:tau_q}). The origin of this logarithmic growth seems to be in
the behavior of the probability distribution function $P(C_4)$ near
$C_4=1$. As shown in the Supporting Information (SI), the numerical data are
consistent with $P(C_4) \propto (1-C_4)^{B(\xi)}$ for $C_4$ close to 1, with
an exponent that grows as $B(\xi(\tw))\sim \log \xi(\tw)$. This behavior of
the correlation function would explain the observed logarithmic growth of
$\tau(q)$. However, just to be on the safe side, we have tried two different
functional forms to fit the numerical data in Fig.~\ref{fig:tau_q}:
\begin{equation}\label{eq:tau-ansatz}
\tau_1(q)=mq\frac{1+c_1q}{1+c_2 q}\,,\quad \tau_2(q)=mq\frac{1+d_1q \log q}{(1+d_2 q)^2}\,.
\end{equation}
Both $\tau_1$ and $\tau_2$ have the same derivative $m$ at $q\!=\!0$. We do
not treat $m$ as a fitting parameter. Rather we take it from the scaling of
the \emph{median} of the distribution $P(C_4)$ with $\xi$ (see SI). Although
both $\tau_1(q)$ and $\tau_2(q)$ make an excellent job at fitting our data
(see again SI), only $\tau_2(q)$ displays the logarithmic growth with $q$, at
large $q$, that we find more plausible.

\section*{Discussion}

Following Ref.~\cite{halsey:86}, we shall discuss our results in terms of a
different stochastic variable, $\alpha=\log C_4(r=\xi(\tw))/\log
[1/\xi(\tw)]$, so that (we drop the argument in $\xi$ for the sake of
shortness)
\begin{equation}\label{eq:f-def}
C_4=\frac{1}{\xi^\alpha}\,,\quad P(C_4) \frac{\mathrm{d} C_4}{\mathrm{d} \alpha}\sim \xi^{f(\alpha)}\,.
\end{equation}
\eqref{eq:f-def} defines the large-deviations function $f(\alpha)$.
Then, we find for the moments of $C_4$
\begin{equation}
\overline{C_4^q}= \int_0^1\text{d}C_4\, P(C_4) C_4^q\sim  \int_0^\infty\text{d}\alpha\, \text{e}^{\log(\xi) [f(\alpha)-q\alpha]}\,.
\end{equation}
For large $\xi$, the above integral is dominated by the maximum of
$[f(\alpha)-q\alpha]$ at some value $\alpha=\alpha^*$:
\begin{equation}
\overline{C_4^q}\sim \frac{1}{\xi^{-[f(\alpha^*)-q\alpha^*]}}\,.
\end{equation}
Comparing with \eqref{eq:tau-q-def}, we realize that
$f(\alpha)$ is just  (minus) the Legendre transform of the singularity spectrum $\tau(q)$:
\begin{equation}\label{eq:Legendre}
    f(\alpha)=-\max_q \big[\tau(q)\, -\, q\alpha \big].
\end{equation}
We show $f(\alpha)$ in Fig.~\ref{fig:Legendre}, as computed from our fitting
ansätze $\tau_1(q)$ and $\tau_2(q)$, in~\eqref{eq:tau-ansatz}. In the range of
Fig.~\ref{fig:tau_q} ---since $\alpha(q)=\tau'(q)$--- the results from the two
ans\"atze can hardly be distinguished. The two, however, differ in that the
range of $\alpha$ for $\tau_2(q)$ goes all the way down to $\alpha=0$ (because
$\alpha_2(q)=\mathrm{d} \tau_2/\mathrm{d} q\sim 1/q$). Indeed, if $\tau(q)$
goes as $\log(q)$ for large $q$, then the large-deviations function goes as
$f(\alpha\to 0)\sim \log(\alpha)$.

\begin{figure}
\centering \includegraphics[width=\linewidth]{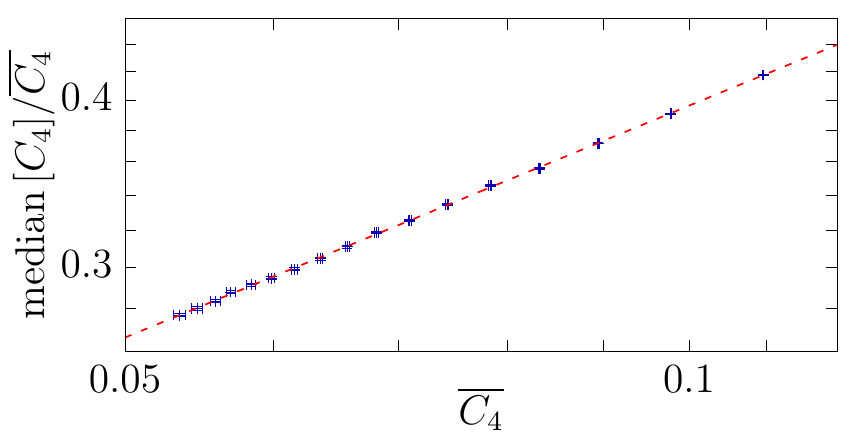}
\caption{Median of the distribution $P[C_4(r=\xi)]$ in units of the first
  moment, $\overline{C_4}(r=\xi)$, versus $\overline{C_4}(r=\xi)$, as computed
  for the spin glass at temperature $T=0.9$. We show data in logarithmic
  scale. Therefore, the dashed line (a power-law fit with exponent $\sim 0.5$,
  see SI for details) appears as a straight line.}
\label{fig:median_vs_C4}
\end{figure}

\begin{figure}[t]
\centering
\includegraphics[width=\linewidth]{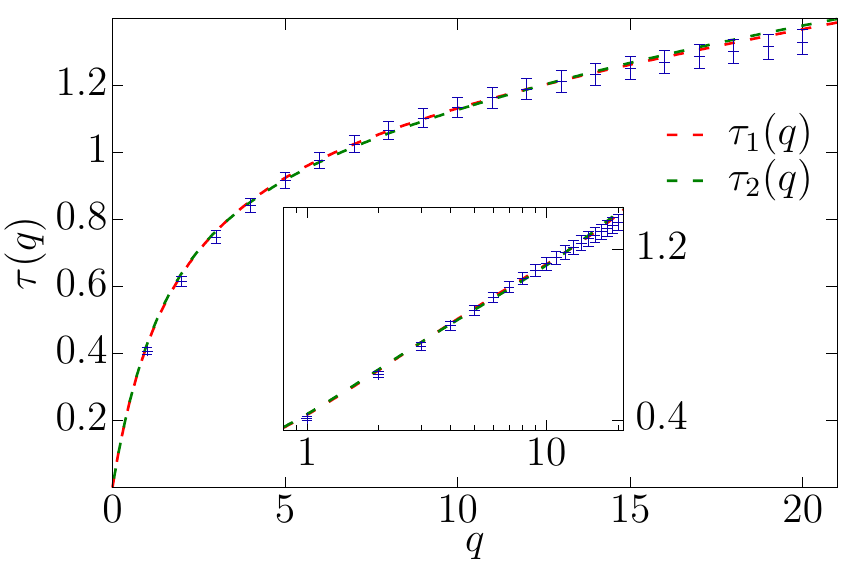}
\caption{Scaling exponent $\tau(q)$ for the $q$-th moment
  $\overline{C_4(r=\xi)^q} \sim \xi^{-\tau(q)}$ computed from simulations of
  the Ising spin glass at $T=0.9$ (see SI for results at $T_\mathrm{c}$). The
  non-linear behavior is a strong indication of multifractality. The dashed
  lines are fits to the functional forms in~\eqref{eq:tau-ansatz} (the
  goodness-of-fit statistics are presented in the SI). The \textbf{inset}
  presents the same data as a function of $\log(q)$.}
\label{fig:tau_q}

\includegraphics[width=\linewidth]{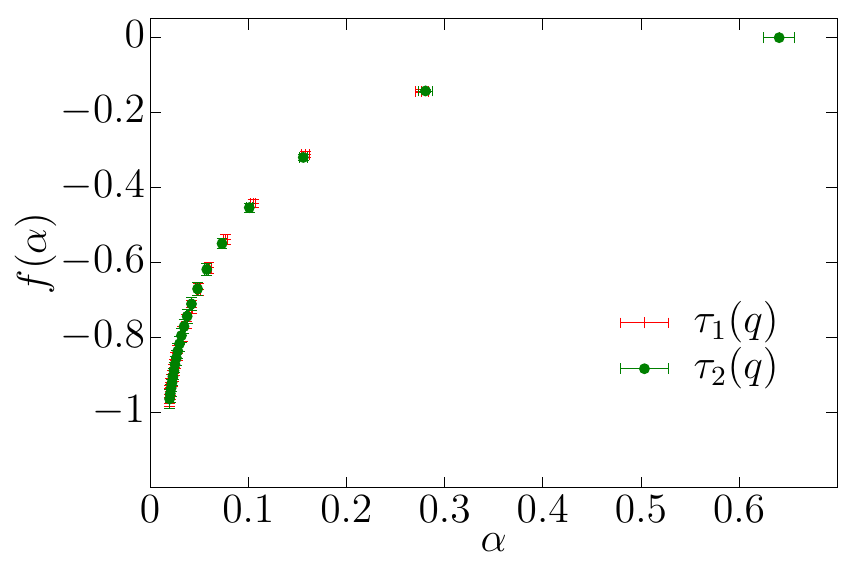}
\caption{Legendre transformation $f(\alpha)$ of function $\tau(q)$,
  see~\eqref{eq:Legendre}, as a function $\alpha=\mathrm{d}\tau/\mathrm{d} q$,
  computed from the fitting ans\"atze in~\eqref{eq:tau-ansatz}. Errors on both
  axes have been obtained as explained in \textbf{Methods} \citep[for further
    details, see][]{yllanes:11}. Since $f(\alpha)$ is the large-deviations
  function of the decay rate $C_4(r=\xi)\sim r^{-\alpha}$, the data show that
  the majority of sites have the median decay rate of approximately $0.65$,
  much larger than the mean decay rate $\alpha \approx 0.4$.}
\label{fig:Legendre}
\end{figure}

Let us recapitulate: the probability of finding a site $\bsym{x}$ with
$C_4(\bsym{x},\bsym{x}+\bsym{r})$ scaling as $1/r^\alpha$ for $r\sim \xi$ goes
in the scaling limit as $\xi^{f(\alpha)}$. There are, hence, a lot more sites
displaying the median scaling exponent $\alpha\approx 0.65$ than there are for
the average scaling $\alpha\approx 0.4$ (because $f(0.65)> f(0.4)$, recall
Fig.~\ref{fig:median_vs_C4}). The larger $\xi(\tw)$ grows, the more pronounced
this difference is. Thus, the expression ``silent majority'' \cite{janus:14c}
could be aptly employed to describe spin-glass dynamics: the central limit
theorem ensures that it is the (somewhat exceptional) average value the one
that can be measured on length scales larger than $\xi(\tw)$ (hence, in
experiments). The experimental-scale dynamics is, however, not completely
blind to these short-scale fluctuations. Indeed, temperature
chaos~\cite{janus:21} ---and, hence, rejuvenation~\cite{janus:23}, which is
certainly experimentally observable~\cite[see, \emph{e.g.},][]{jonason:98}---
is ruled by statistical fluctuations at the scale of $r$ smaller than, or
similar to, $\xi(\tw)$.

Our data show that varying $T$ simply changes $\tau(q)$ by an essentially
constant factor [\emph{e.g.}, $\tau(q,T_\mathrm{c})\approx 1.5 \tau(q,T=0.9)$,
  see SI]. Furthermore, Fig.~\ref{fig:C4_norm_vs_C4} makes us confident that,
taking $\overline{C_4}$ as scaling variable instead of $\xi(\tw)$, the overall
picture is essentially temperature independent for $T<T_{\mathrm{c}}$.

Whether or not multifractal behavior is also present in equilibrium
correlation functions in the spin-glass phase stands out as an interesting
open question. Statics-dynamics
equivalence~\cite{franz:98,janus:10b,wittmann:16,janus:17} suggests that the
answer will be positive.

As a final remark, let us stress that ongoing efforts to build a
mathematically rigorous theory of non-equilibrium spin-glass dynamics through
the concept of the maturation metastate (see~\citealp{jensen:21} and
references therein) should take into account the extreme spatial heterogeneity
unveiled in this work.

\section*{Methods}

\subsection*{Model and simulations} We focus on the Edwards-Anderson model (EA) in a simple cubic lattice  with linear size $L=160$ and periodic boundary conditions. Our $S_{\bsym{x}}=\pm 1$ spin, placed at the lattice sites, interact with their nearest neighbors through the Hamiltonian:
\begin{equation}\label{eq:Hamiltonian}
\mathcal{H}= - \sum_{\langle \bsym{x},\bsym{y}\rangle} J_{\bsym{x},\bsym{y}}
S_{\bsym{x}}S_{\bsym{y}}\, .
\end{equation}
The coupling constants $J_{\bsym{x},\bsym{y}}$ are independent random
variables ($J_{\bsym{x},\bsym{y}}=\pm 1$ with equal probability), fixed once
and for all at the beginning of the simulation (this is named quenched
disorder). A realization of the couplings is called a \emph{sample}. We shall
use 16 samples in this work.  In general, errors will be computed with a
jackknife method over the samples \citep[see, for
  instance,][]{amit:05,yllanes:11}. We have also considered a diluted Ising
model (see below), as a baseline model displaying domain-growth
off-equilibrium dynamics.

We have simulated the model in~\eqref{eq:Hamiltonian} through a Metropolis
dynamics on the Janus II supercomputer~\cite{janus:14}. Our time unit is a
full-lattice sweep, roughly equivalent to a picosecond of physical
time~\cite{mydosh:93}. The critical temperature for this model is
$T_\mathrm{c}=1.1019(29)$~\cite{janus:13}.

For each sample, we have simulated $\NR=512$ statistically independent system
copies or \emph{replicas}. We denote by $\langle \cdots\rangle$ the average
over thermal noise for one sample (as explained below, we obtain unbiased
estimators of the thermal expectation values $\langle \cdots\rangle$ by
averaging over the replicas). The subsequent average over samples is denoted
by an overline ($\overline{\langle\cdots\rangle}$).

The main quantity of interest is the correlation function 
\begin{equation}\label{eq:C4-x-y-tw}
    C_4(\bsym{x},\bsym{y},\tw) = \langle S_{\bsym{x}} (\tw)  S_{\bsym{y}}(\tw)\rangle^2\,.
\end{equation}

Note that, for a given sample and $(\bsym{x},\bsym{y},\tw)$,
$C_4(\bsym{x},\bsym{y},\tw)$ is not a stochastic variable. However, it
\emph{is} a stochastic variable if we regard the variations induced by the
choice of couplings $J_{\bsym{x},\bsym{y}}$ and over the considered sites
$(\bsym{x},\bsym{y},\tw)$. We term these stochastic variables $C_4$, without
arguments.

As explained in the next paragraph, although $C_4(\bsym{x},\bsym{y},\tw)$
cannot be computed with a finite number of replicas, unbiased estimators of
its moments can be computed. In particular, previous work has mostly focused
on the average correlation function
\begin{equation}\label{eq:C4-av}
    C_4^{\text{av}}(\bsym{r},\tw)=\frac{1}{L^3}\sum_{\bsym{x}}
    \overline{C_4(\bsym{x},\bsym{y}=\bsym{x}+\bsym{r},\tw)}\,.
\end{equation}
Cubic symmetry, present in averages over the samples, allows us to average
over the three equivalent displacements $\bsym{r}=(r,0,0)$ and
permutations. We shall use the shorthand $C_4^{\text{av}}(r,\tw)$ to indicate
this average over the three equivalent $\bsym{r}$. To compute the coherence
length $\xi(\tw)$ we follow~\cite{janus:08b,janus:09b,janus:18} and compute
the integrals
\begin{equation}
I_n(\tw)= \int_0^\infty r^n\, C^{\text{av}}_4(r,\tw)\mathrm{d}r\,.    
\end{equation}
Then, $\xi(\tw)=I_2(\tw)/I_1(\tw)$. 

As stated above, we have simulated, as a null experiment, the link-diluted
Ising model (DIL). The only difference with the Hamiltonian
in~\eqref{eq:Hamiltonian} is in the choice of the couplings:
$J_{\bsym{x},\bsym{y}}=1$ (with $70\%$ probability) or
$J_{\bsym{x},\bsym{y}}=0$ (with $30\%$ probability). Since all couplings are
positive or zero, this is a ferromagnetic system without frustration.  All our
simulation and analysis procedures are identical for the DIL and EA models.
The critical temperature is
$T^{\text{DIL}}_{\mathrm{c}}=3.0609(5)$~\cite{berche:04}. Actually, this is
twice the value reported in~\cite{berche:04} due to our use of an Ising,
rather than Potts, formulation.  In fact, with some abuse of language, in the
main text we refer to DIL temperatures as $\tilde{T}=0.9$, $\tilde{T}=0.8$ or
$0.7$ rather than to their real value $T^{\mathrm{DIL}}=\tilde{T}\,
(T_\mathrm{c}^{\text{DIL}}/T_\mathrm{c}^{\text{EA}})$.

The range of coherence length and simulation times in this study can be found in Table~\ref{tab:simulations}.

\begin{table}
    \centering
\caption{Maximum $\tw$ and coherence length reached for each of our models and simulation temperatures.}
    \begin{tabular}{ccccc}
         $T$ or $\tilde{T}$& $\tw$(EA) & $\xi_{\mathrm{max}}$(EA)& $\tw$(DIL)
      & $\xi_{\mathrm{max}}$(DIL)\\ \hline 
         0.7 & 46531866276 & 12 & 498 & 15 \\
         0.8 & 18734780191 & 15 & 919 & 21 \\
         0.9 & 15172184825 & 20 & 954 & 23 \\\hline 
    \end{tabular}
    \label{tab:simulations}
\end{table}

\subsection*{Unbiased estimators of powers of \boldmath
  $C_4(\bsym{x},\bsym{y},\tw)$}
Given $\bsym{x}$ and $\bsym{y}$, we need an
unbiased estimator of $C_4^q(\bsym{x},\bsym{y},\tw)=\langle S_\bsym{x}(\tw)
S_\bsym{y}(\tw)\rangle^{2q}$.  Note that the $q=1$ instance is needed to
evaluate~\eqref{eq:C4-av}.

Should we have (at least) $2q$ replicas at our disposal, a tentative solution
would be provided by the estimator
\begin{equation}\label{eq:C4-q-poor}
[C_4(\bsym{x},\bsym{y},\tw)]_q^\text{poor} =\prod_{a=1}^{2q}\, S^{(a)}_\bsym{x}(\tw) S^{(a)}_\bsym{y}(\tw)\,.
\end{equation}
$[C_4(\bsym{x},\bsym{y},\tw)]_q^\text{poor}=(-1)^p$, where $p$ is the number
of replicas for which $S^{(a)}_\bsym{x}(\tw) S^{(a)}_\bsym{y}(\tw)=-1$.
However, the statistical independence of the different replicas ensures for
the expectation value $\langle
[C_4(\bsym{x},\bsym{y},\tw)]_q^\text{poor}\rangle = \langle S_\bsym{x}(\tw)
S_\bsym{y}(\tw)\rangle^{2q}$.

Nevertheless, if we have at our disposal a number of replicas $\NR\gg 2q$, as
is our case, the solution in~\eqref{eq:C4-q-poor} is very
unsatisfactory. Rather, one would like to consider all possible picks of $2q$
\emph{different} replicas (out of the $\NR$ possible choices), compute
$[C_4(\bsym{x},\bsym{y},\tw)]_q^\text{poor}$ for every pick, and take the
average of those products.

To achieve our goal, we have solved the following auxiliary combinatorial
problem. Given a set of $\NR$ different signs $c_a=\pm 1$, $M$ of which
negative, we have computed $\tilde P(\NR,M;S,p)$, namely the probability of
getting $p$ negative signs in a pick (with uniform probability) of $S$
\emph{distinct} signs. With this probability in our hands, the solution is
straightforward. We just need to look at our set $S^{(a)}_\bsym{x}(\tw)
S^{(a)}_\bsym{y}(\tw)$, $a=1,2\ldots,\NR$, count the number $M$ of them that
turn out to be negative and compute the estimator
\begin{equation}\label{eq:C4-q-good}
[C_4(\bsym{x},\bsym{y},\tw)]_q = G(\NR,M,q)\,,
\end{equation}
\begin{equation}
  G(\NR,M,q)=\sum_{p=0}^{2q}\, (-1)^p\,  \tilde P(\NR,M;
  S=2q,p)\,.\label{eq:G-def}
\end{equation}

$[C_4(\bsym{x},\bsym{y},\tw)]_q$ is an unbiased estimator of $\langle
S_\bsym{x}(\tw) S_\bsym{y}(\tw)\rangle^{2q}$, because it is an average over
all possible (poor, but unbiased) estimators in~\eqref{eq:C4-q-poor}. Our
computation of $\tilde P(\NR,M; S=2q,p)$ is explained in the SI.

\subsection*{The probability distribution of the correlation function}
We wish to study the probability distribution function (pdf) for $\langle
S_\bsym{x}(\tw) S_{\bsym{x}+\bsym{r}}(\tw)\rangle^2$ (periodic boundary
conditions are assumed for $\bsym{x}+\bsym{r}$). We have only considered
displacements $\bsym{r}=(r,0,0)$ ---and permutations--- and we have chosen the
measuring times in such a way that $r=\xi(\tw)$.

Note that, given the starting point $\bsym{x}$ and the sample
$\{J_{\bsym{x},\bsym{y}}\}$, $\langle S_\bsym{x}
S_{\bsym{x}+\bsym{r}}\rangle^2$ is not a fluctuating quantity. Hence, we are
referring to the pdf as $\bsym{x}$ and the sample vary. $\langle S_\bsym{x}
S_{\bsym{x}+\bsym{r}}\rangle^2$ can be computed exactly only in the limit
$\NR\to\infty$. However, as explained in the previous paragraph, we can
compute without bias its $q$-th moment provided that the number of replicas at
our disposal is $\NR\geq 2q$.

The basic object we compute from our simulation is the pdf ${\cal
  P}(M;\NR,\xi)$, namely the probability, as computed over the starting point
$\bsym{x}$ and the samples, that exactly $M$ of the $\NR$ signs
$S^{(a)}_\bsym{x}(\tw) S^{(a)}_{\bsym{x}+\bsym{r}}(\tw)$ turn out to be $-1$
in our simulation of this specific sample. Hence, the unbiased estimator of
the $q$-th moment of $\langle S_\bsym{x}(\tw)
S_{\bsym{x}+\bsym{r}(\tw)}\rangle^2$ with $r=\xi(\tw)$ is
\begin{equation}
    \overline{C_4^q}(\xi)=\sum_{M=0}^{\NR} {\cal P}(M;\NR,\xi) G(\NR,M,q)\,,
\end{equation}
where $G(\NR,M,q)$ was defined in~\eqref{eq:G-def}.

Unfortunately, the median of the pdf for $\langle S_\bsym{x}(\tw)
S_{\bsym{x}+\bsym{r}}(\tw)\rangle^2$ is more difficult to compute. Our
strategy, explained in full detail in the SI, consists in computing biased
estimators of the median, with bias of order $1/\NR$. Then we compute these
biased estimators for a sequence $\NR'=32, 64, 128, 256$ and $512$, and
proceed to an extrapolation $\NR\to\infty$. We obtain the ${\cal
  P}(M';\NR',\xi)$ from their $\NR=512$ counterpart as
\begin{eqnarray}
  {\cal P}(M';\NR',\xi)&=&
  \sum_{M=0}^{\NR} {\cal P}(M;\NR,\xi)\,\times \nonumber\\
  &\times & P(\NR,M;S=\NR',p=M')\,.
\end{eqnarray}
The probabilities $P(\NR,M;S,p)$ were defined in the previous subsection in this \textbf{Methods} section.

\subsection*{Computation of \boldmath $\tau(q)$} In order to minimize corrections to scaling, we have fitted the normalized moments as
\begin{equation}
\frac{\overline{C_4^q}}{\overline{C_4}^q}=\Big[\frac{A_q}{\overline{C_4}}\Big]^{\beta(q)}\,,\quad
\tau(q)=\tau(1) [q-\beta(q)]\,.
\end{equation}
Fig.~\ref{fig:C4_norm_vs_C4} provides an example. In order to obtain good
fits, we have needed to discard (at most) one data point corresponding to the
smallest $\xi(\tw)$. An advantage of this method is that we only need to
consider the $\xi(\tw)$ dependence to obtain $\tau(1)$, as shown in
Fig.~\ref{fig:C4_vs_xi}. The full procedure is illustrated in the SI.

To compute errors we have followed the strategy of~\cite{yllanes:11}, namely
carrying out all fits separately for each jackknife block (when minimizing
$\chi^2$ to perform the fits, we only consider the diagonal elements of the
covariance matrix). Errors in the fit parameters are obtained from the
fluctuations of the jackknife blocks.

\subsection*{Computation of \boldmath $M(\bsym{x},\bsym{r})$} 
The order-of-magnitude factor in \eqref{eq:M-def} is computed as $\log |
[C_4(\bsym{x},\bsym{x}+\bsym{r},\tw)]_1 |/\log
C_4^{\text{av}}(\sqrt{r_x^2+r_y^2+r_z^2})$, where
$[C_4(\bsym{x},\bsym{x}+\bsym{r},\tw)]_1$ is the $q=1$ estimator
in~\eqref{eq:C4-q-good} (as computed with $\NR=512$). $C_4^\text{av}(r)$ is
interpolated to non-integer arguments using a fit obtained from data with
integer $r$ (see SI).

\section*{Acknowledgments}

\begin{acknowledgments} 
We thank Roberto Benzi and Luca Biferale for relevant discussions. This work
was supported in part by Grants No.PID2022-136374NB-C21, PID2020-112936GB-I00,
PID2019-103939RB-I00, PGC2018-094684-B-C21, PGC2018-094684-B-C22 and
PID2021-125506NA-I00,funded by MCIN/AEI/10.13039/501100011033 by “ERDF A way
of making Europe” and by the European Union. The research has received
financial support from the Simons Foundation (grant No.~454949, G. Parisi) and
ICSC – Italian Research Center on High Performance Computing, Big Data and
Quantum Computing, funded by European Union – NextGenerationEU. DY was
supported by the Chan Zuckerberg Biohub. IGAP was supported by the Ministerio
de Ciencia, Innovaci\'on y Universidades (MCIU, Spain) through FPU grant
No. FPU18/02665. JMG was supported by the Ministerio de
Universidades and the European Union NextGeneration EU/PRTR through 2021-2023
Margarita Salas grant. IP was supported by LazioInnova-Regione Lazio under the
program Gruppi di ricerca2020 - POR FESR Lazio 2014-2020, Project NanoProbe
(Application code A0375-2020-36761).
\end{acknowledgments}

\appendix

\begin{center}
  \bf SUPPORTING INFORMATION
\end{center}

\section{The auxiliary combinatorial problem}

\subsection{The computation of $\tilde P(\NR,M;S,p)$}
As stated in the main text, we need to consider the following problem: given a
set of $\NR$ different signs $c_a=\pm 1$, $M$ of which negative, we need to
obtain $\tilde P(\NR,M;S,p)$, namely the probability of getting $p$ negative
signs in a pick (with uniform probability) of $S$ \emph{distinct} signs.  In
order to organize the computation, we shall consider both the sign labels and
the pick orderings as distinguishable. Hence, the number of possible picks is
\begin{equation}
    N_{\text{picks}}=\frac{\NR!}{(\NR-S)!}\,.
\end{equation}
Let us denote by $K(\NR,M;S,p)$ the number of picks of $S$ distinct signs that
contain exactly $p$ negative signs. Hence,
\begin{equation}
 \tilde P(\NR,M;S,p)=\frac{K(\NR,M;S,p)}{N_{\text{picks}}}\,.
\end{equation}
$K(\NR,M;S,p)$ can be obtained as a product of three factors

\begin{equation}
  K(\NR,M;S,p)=F_1F_2F_3\,,
\end{equation}
where
\begin{eqnarray}
  F_1&=& \frac{M!}{p!(M-p)!}\,,\nonumber\\
  F_2&=&\frac{S!}{(S-p)!}\,,\\
  F_3&=&\frac{(\NR-M)!}{(\NR-M-S+p)!}\,.\nonumber
\end{eqnarray}
In the above expression, whenever the factorial of a negative integer arises
it should interpreted as $\infty$ (hence the corresponding factor
vanishes). The meaning of the different factors is as follows:
\begin{itemize}
    \item $F_1$ is the number of ways that we can choose $p$ tags of negative
      signs among $M$ possibilities.
    \item $F_2$ is the number of ways in which a given set of $p$ tags of
      negative signs can be extracted: the first tag can be obtained in the
      first selection, or in the second, etc. So there are $S$ possibilities
      for the first tag, which leaves us with $S-1$ possibilities for the
      second tag, and so on.
    \item Finally, $F_3$ is concerned with the $S-p$ positive signs that we
      need to complete the pick of $S$ signs. We have $\NR-M$ choices for the
      first tag to be chosen, $\NR-M-1$ for the second tag, and so forth.
\end{itemize} 

We have found it preferable, however, to make use of a Pascal-Tartaglia-like
relation. To obtain it, it is useful to think of a pick of $S$ signs as two
consecutive picks. We get $S-1$ signs from the first pick, and the last sign
$c_\alpha$ is chosen only afterwards:
\begin{eqnarray}
  K(\NR,M;S,p)&=& {\cal A} \, K(\NR,M;S-1,p-1) \nonumber\\
  &+&{\cal B} \, K(\NR,M;S-1,p),
\end{eqnarray}
where ${\cal A}$ is the number of negative signs available for the last pick
(given that we obtained $p-1$ negative signs from the first pick), while
${\cal B}$ is the number of positive signs available for the last pick (given
that we already obtained $p$ negative signs from the first pick),
specifically:
\begin{equation}
  {\cal A}=\text{max}\{0,M-p+1\}\,,
\end{equation}
\begin{equation}
  {\cal B}=\text{max}\{\,0\,,\, \NR-
    (S-1)\, -\, \text{max}\{0,M-p\}\,\}\,.
\end{equation}
The recursion relation for $K(\NR,M;S,p)$ instantaneously translates to a
recursion relation for the probability:
\begin{eqnarray}
\tilde P(\NR,M;S,p)&\!=\!& \frac{{\cal A}}{\NR-S+1} \, \tilde P(\NR,M;S-1,p-1) \nonumber\\
&\!+\!&\frac{{\cal B}}{\NR-S+1} \, \tilde P(\NR,M;S-1,p).
\end{eqnarray}
Starting the recursion from
\begin{eqnarray}
  P(\NR,M;S=1,p=0)&=&\frac{\NR-M}{\NR}\,,\nonumber\\
  P(\NR,M;S=1,p=1)&=&\frac{M}{\NR}\,,
\end{eqnarray}
we can simply and accurately compute $P(\NR,M;S,p)$ for whatever values of
$M$, $S$ and $p$ we need, respecting, of course, the obvious bounds:
\begin{equation}
\NR\geq M,S,p\,,\quad M\geq p\,,\quad S\geq p\,.
\end{equation}

\subsection{A useful symmetry} The probability that we have computed in the previous paragraph presents a \emph{spin-flip} symmetry
\begin{equation}\label{eq:S-flip}
    \tilde P(\NR,M; S,p)=\tilde P(\NR,\NR-M; S,S-p)\,.
\end{equation}
The simplest way to show that the symmetry is present is noticing that the map
of the signs set $\{c_a\}_{a=1}^{\NR}$ into $\{ -c_a\}_{a=1}^{\NR}$ is
bijective, transforming $M$ into $\NR-M$ and a pick of $p$ negative signs into
a pick of $S-p$ negative signs.

\eqref{eq:S-flip} implies a symmetry in the quantity we use to estimate the
moments of the correlation function, recall the main text,
\begin{eqnarray}\label{eq:C4-q-good-SI}
[C_4(\bsym{x},\bsym{y},\tw)]_q &=&
G(\NR,M,q)\,,\\
G(\NR,M,q)&=&\sum_{p=0}^{2q} (-1)^p\, \tilde P(\NR,M;
S\!=\!2q,p).\label{eq:G-def-SI}
\end{eqnarray}
Hence, the symmetry in ~\eqref{eq:S-flip} implies
\begin{equation}\label{eq:symmetry}
    G(\NR,M,q)=G(\NR,\NR-M,q)\,.
\end{equation}
This consideration is important when we aim to compute the median of the
stochastic variable $C_4$ (remember that we only have in our hands estimators
such as $[C_4]_{q=1}$) A moment's thought reveals that, for even $\NR$, the
minimum value of $G(\NR,M,q=1)$ is reached at $M=\NR/2$. Indeed,
\begin{equation}
G(\NR,M,q=1)=\frac{(\NR-2M)^2 -\NR}{\NR(\NR-1)}\, .
\end{equation}
Hence, one may like to use a symmetrized probability, rather than the
probability discussed in the main text, ${\cal P}(M;\NR,\xi)$, (the
probability, as computed over the starting point $\bsym{x}$ and the samples,
that exactly $M$ of the $\NR$ signs $S^{(a)}_\bsym{x}(\tw)
S^{(a)}_{\bsym{x}+\bsym{r}}(\tw)$ turn out to be $-1$ in our simulation of
this specific sample)~\footnote{Recall that we set times such that
  $r=\xi(\tw)$.}. Specifically,
\begin{eqnarray}
{\cal P}_s\left(n=\frac{\NR}{2};\NR,\xi\right)&=&{\cal P}(n;\NR,\xi)\,,\\
{\cal P}_s\left(\frac{\NR}{2}< n\leq \NR;\NR,\xi\right)&=&{\cal
  P}_s(n;\NR,\xi)\nonumber\\
&+&{\cal P}_s(\NR -n;\NR,\xi)\,.\nonumber
\end{eqnarray}
Hence, the moments of $C_4$ can be computed as
\begin{equation}
    \overline{C_4^q}(\xi)=\sum_{n=\NR/2}^{\NR} {\cal P}_s(n;\NR,\xi)
    G(\NR,n,q)\,.
\end{equation}
It is then clear that we need to estimate medians from the symmetrized
probability ${\cal P}_s(n;\NR,\xi)$, rather than from ${\cal P}(M;\NR,\xi)$.

\section{The computation of the medians}
\subsection{Computing the medians from our biased estimators}
Our analysis starts from the symmetrized probability ${\cal
  P}_s(n;\NR,\xi)$. We begin by computing the cumulative distribution
\begin{equation}
{\cal S}(n) = \sum_{\ell=\NR/2}^{n} {\cal P}_s(\ell;\NR,\xi)\,,
\end{equation}
and determine the smallest integer $n^*\geq \NR/2$ such that ${\cal S}(n^*) >
0.5$. Indeed, we assign positive weights $\omega_-$ and $\omega_+$ to
$(n^*-1)$ and $n^*$, in such a way that
\begin{equation}
   \omega_-+\omega_+=1\,, \quad \omega_-{\cal
     S}(n^*-1)\ +\ \omega_+{\cal S}(n^*)=0.5\,.
\end{equation}
In other words we linearly interpolate ${\cal S}(n)$ in $n$, in order to
obtain a cumulative exactly equal to 0.5.

However, we need to translate the integer $n^*$ into a real number, namely an
estimator of the median of $C_4$. In order to do that, we need to take one
step back and consider the physical meaning of the fact that exactly $M$ of
the $\NR$ signs $S^{(a)}_\bsym{x}(\tw) S^{(a)}_{\bsym{y}}(\tw)$ turn out to be
$-1$ in our simulation of an specific sample. Indeed, one immediately notices
that
\begin{eqnarray}
\frac{\NR-2M}{\NR}&=&\frac{1}{\NR}\sum_{a=1}^\NR\, S^{(a)}_\bsym{x}(\tw) S^{(a)}_{\bsym{y}}(\tw) \\
&=& \langle S_\bsym{x}(\tw) S_{\bsym{y}}(\tw)\rangle +
\eta\sqrt{\frac{1-C_4}{\NR}}\,,\nonumber
\end{eqnarray}
where $C_4=\langle S_\bsym{x}(\tw) S_{\bsym{y}}(\tw)\rangle^2$ and $\eta$ is a
random variable that verifies
\begin{equation}
\langle \eta\rangle=0\,,\quad \langle \eta^2\rangle=1\,.
\end{equation}
Furthermore, $\eta$ is statistically uncorrelated with $\langle
S_\bsym{x}(\tw) S_{\bsym{y}}(\tw)\rangle$ and, because of the central limit
theorem, tends to a normally distributed stochastic variable in the limit of
large $\NR$. Hence, we may consider the (very biased) estimator of $C_4$
($\text{sgn}(x)=x/|x|$):
\begin{eqnarray}\label{eq:SxSy2-DISCUSSION}
[S_\bsym{x}(\tw) S_{\bsym{y}}(\tw)]^2 &\!\equiv\! &
\left(\frac{\NR-2M}{\NR}\right)^2 \\
= C_4 &\!+\!& 2\eta \,\text{sgn}(\langle S_\bsym{x}(\tw)
S_{\bsym{y}}(\tw)\rangle)\sqrt{\frac{C_4(1-C_4)}{\NR}}\nonumber\\
&\!+\!&\eta^2
\frac{(1-C4)}{\NR}\,.\nonumber
\end{eqnarray}
We do not expect problems from the term linear in $\eta$. Indeed for $\NR$
large enough, $\eta$ approaches a normal variable (so, symmetrically
distributed around zero), hence it should not cause bias on the estimation of
the median. The real problem comes from the term proportional to $\eta^2$,
inducing departures from the true value of $C_4$ of order $1/\NR$.  Hence we
may consider our first estimator of the median
\begin{eqnarray}\label{eq:mediana_SxSy_2}
\text{median}([S_\bsym{x}(\tw)
  S_{\bsym{y}}(\tw)]^2)&=&\omega_-\!\left(\frac{\NR-2
  (n^*-1)}{\NR}\right)^2\nonumber\\
&+& \omega_+\! \left(\frac{\NR-2n^*}{\NR}\right)^2.
\end{eqnarray}
As explained above, we expect that this estimator of the median of $C_4$ will
have a bias of order $1/\NR$.

Alternatively, we may consider an unbiased estimator of $C_4$:
\begin{eqnarray}\label{eq:C4-DISCUSSION}
  [C_4]&=&G(\NR,M,q=1)= \frac{(\NR-2M)^2 -\NR}{\NR(\NR-1)}\\
  &=&C_4+ 2\eta
\,\text{sgn}(\langle S_\bsym{x}(\tw) S_{\bsym{y}}(\tw)\rangle)\sqrt{\frac{\NR
    C_4(1-C_4)}{(\NR-1)^2}} \nonumber\\
&+&\frac{(\eta^2-1)}{\NR-1}\,.\nonumber
\end{eqnarray}
Notice that, although the expectation value of $\eta^2-1$ is zero, this term
is not symmetrically distributed around zero (not even in the limit
$\NR\to\infty$, when $\eta$ is normally distributed). Hence the $\eta^2-1$
will also distort the computation of the median by a quantity of order
$1/\NR$. Accordingly, we expect corrections of order $1/\NR$ for the
corresponding estimator of the bias of the median of $C_4$
\begin{eqnarray}\label{eq:mediana_C4}
  \text{median}([C_4])&=&\omega_-G(\NR,n^*-1,q=1)\nonumber\\
  &+&\ \omega_+G(\NR,n^*,q=1)\,.
\end{eqnarray}

\begin{figure}
\centering \includegraphics[width=\linewidth]{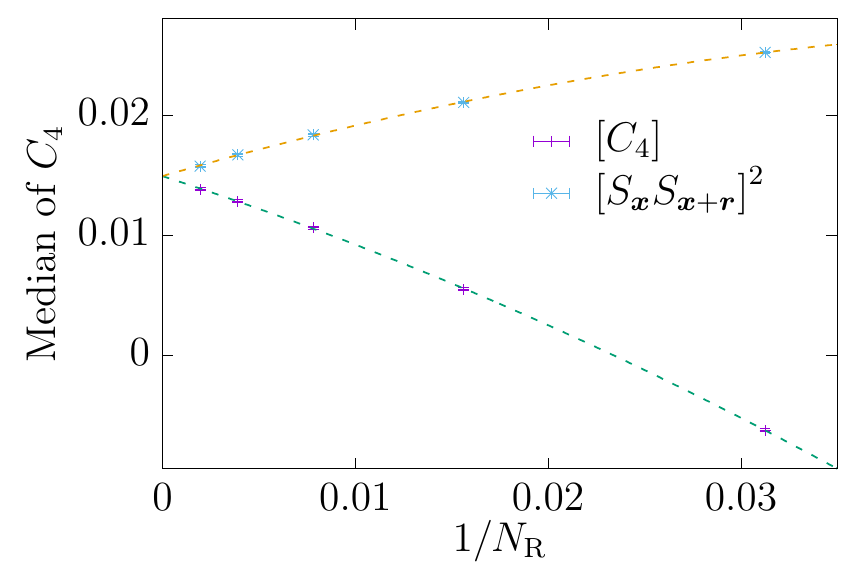}
\caption{Biased estimators of the median, see~\eqref{eq:SxSy2-DISCUSSION}
  and~\eqref{eq:C4-DISCUSSION}, as functions of the inverse number of
  replicas, $\NR$, for $T=0.9$ and $\xi(\tw)=20$. The dashed lines are fits to
  quadratic functions in order to extrapolate to $\NR\rightarrow\infty$
  (complete fit statistics are available in Table~\ref{tab:fit_summary}).}
\label{fig:mediana_vs_NR}
\end{figure}

The extrapolation to $\NR\to\infty$ from both biased estimators is illustrated
in Fig.~\ref{fig:mediana_vs_NR}.

\begin{figure}
\centering \includegraphics[width=\linewidth]{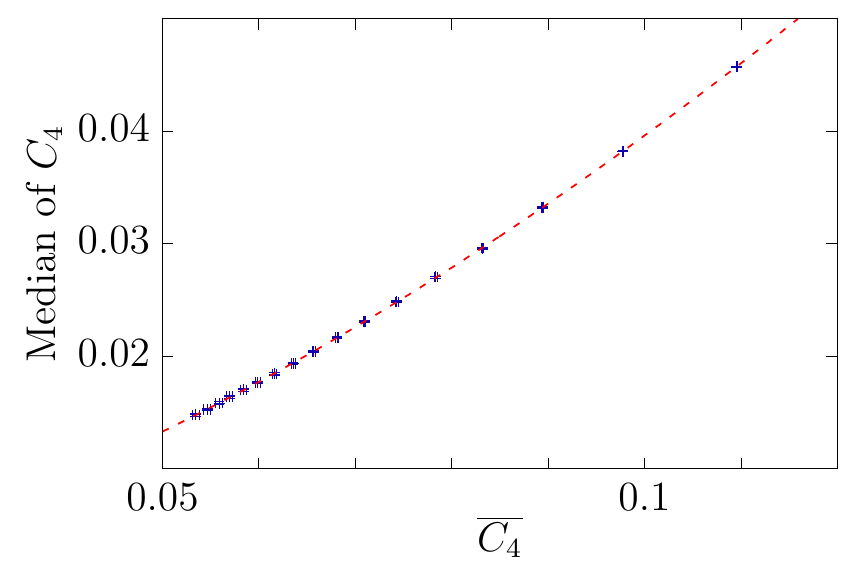}
\caption{Median of the distribution $P[C_4(r=\xi)]$ versus the first moment,
  $\overline{C_4}(r=\xi)$, for $T=0.9$. The median is calculated through the
  extrapolation to $\NR\rightarrow\infty$ shown in
  Fig.~\ref{fig:mediana_vs_NR}. The dashed line is a fit to a power law in
  order to obtain the slope of $\tau(q)$ near $q=0$ (complete fit information
  available in Table~\ref{tab:fit_summary}).}
\label{fig:mediana_vs_C4}
\end{figure}

\subsection{The probability distribution near \boldmath $C_4=1$}

As we have shown in the main text, the scaling exponent for the $q$-th moment
$\tau(q)$, \emph{i.e.}, $\overline{C_4^q}\sim 1/[\xi(\tw)]^{\tau(q)}$, goes as
$\tau(q)\sim\log q$ for large $q$. This is to be expected if the probability
distribution function for $C_4$ near 1 behaves as
\begin{equation}
    P(C_4\to 1) \propto (1-C_4)^{B(\xi)}\,,
\end{equation}
with an exponent $B(\xi)$ that grows logarithmically with $\xi$. Indeed, a
simple saddle-point estimation yields
\begin{eqnarray}
  \overline{C_4^q}&=&\int_0^1\mathrm{d}C_4\, P(C_4) C_4^q\\
  &\sim & 
    \int_0^1\mathrm{d}C_4\, \exp{[B(\log(1-C_4) +q \log C_4]}\approx
    \frac{1}{q^B}\,.\nonumber
\end{eqnarray}
Hence, if $B(\xi)\sim A \log(\xi)$ ($A$ is an amplitude)
\begin{equation}
\frac{1}{q^{A\log\xi}}=\frac{1}{\xi^{A\log q}}\,.
\end{equation}

\begin{figure}
\centering \includegraphics[width=\linewidth]{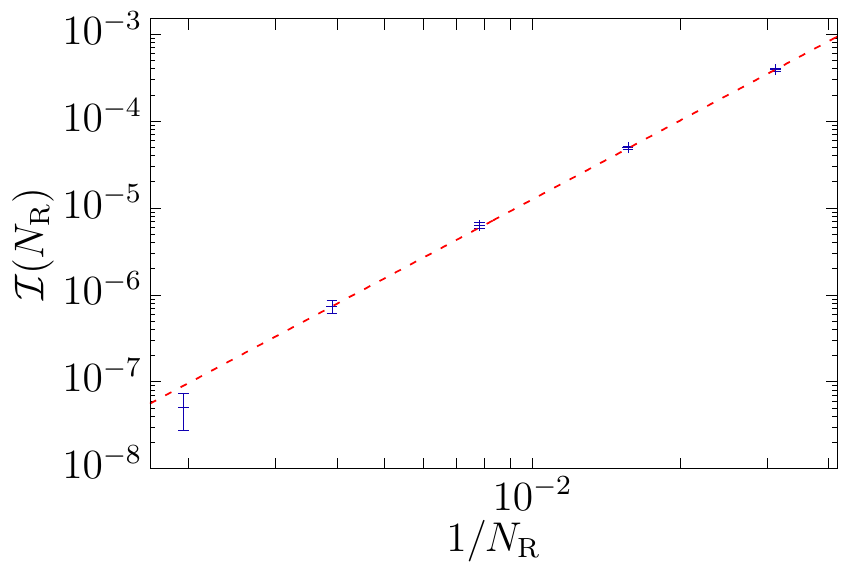}
\caption{Expectation value ${\cal I}(\NR)$, see
  ~\eqref{eq:valor_esperado_exp}, versus the inverse of the number of replicas
  $\NR$, at $T=0.9$ and $\xi(\tw)=20$. The dashed line is a fit
  to~\eqref{eq:fit_valor_esperado_exp} with $\NR\in [64,512]$ (full details of
  the fit available in Table~\ref{tab:fit_summary}). }
\label{fig:exp_vs_NR}
\end{figure}

\begin{figure}
\centering \includegraphics[width=\linewidth]{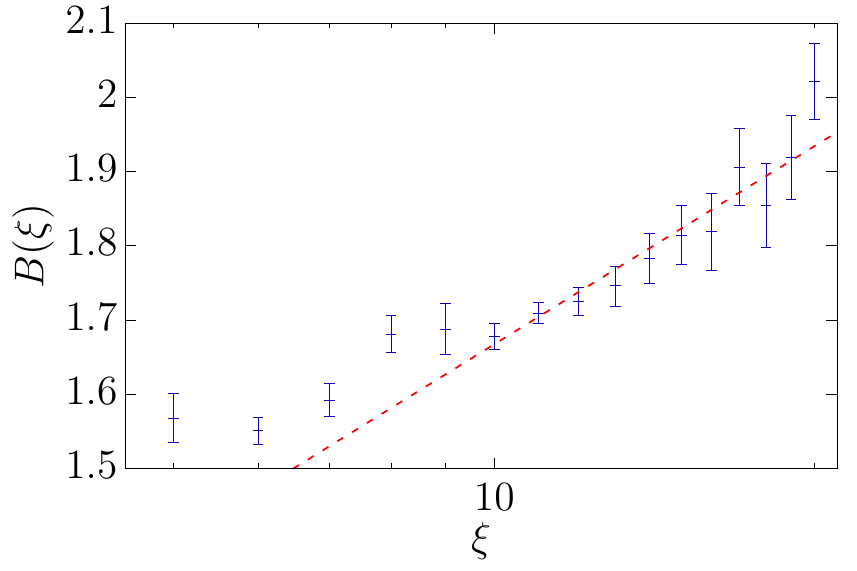}
\caption{Exponent $B(\xi)$ for $T=0.9$. The dashed line is a fit to
  $a+b\log(\xi)$ for $\xi\geq10$ (see Table~\ref{tab:fit_summary} for complete
  fit statistics). These results make more plausible a logarithmic behavior of
  $\tau(q)$ for large $q$.}
\label{fig:exp_vs_xi}
\end{figure}

Equipped with the above intuition, we have checked the $P(C_4\to 1)$ as
computed from the $\NR=512$ estimator in~\eqref{eq:C4-DISCUSSION}. We obtain a
good fit to $(1-C_4)^{B(\xi)}$, but the determination of the exponent $B(\xi)$
is quite difficult, as it depends significantly on the fitting range. This is
why we have turned to a different strategy. We have considered the following
expectation value
\begin{equation}\label{eq:valor_esperado_exp}
    {\cal I}(\NR)=\sum_{n=\NR/2}^{\NR} {\cal P}_s(n;\NR,\xi)
    \mathrm{e}^{-4(\NR-n)}\,.
\end{equation}
The above sum is clearly dominated by values of $n$ near $\NR$ where,
recall~\eqref{eq:SxSy2-DISCUSSION}, $C_4\approx 1- 4 (\NR-n)/\NR$. Hence, we
can approximate
\begin{eqnarray}\label{eq:fit_valor_esperado_exp}
{\cal I}(\NR) &\approx & \int_0^1\!\mathrm{d}C_4\, P(C_4)
\mathrm{e}^{-\NR(1-C_4)}\\
&\sim &\!\int_0^1\!\mathrm{d}C_4\, (1-C_4)^{B(\xi)}
\mathrm{e}^{-\NR(1-C_4)}\sim\frac{1}{\NR^{1+B(\xi)}}\,.\nonumber
\end{eqnarray}
Then, our chosen strategy has been to fit our data for ${\cal I}(\NR)$ as a
power law in $1/\NR$, see Fig.~\ref{fig:exp_vs_NR}. Generally speaking, we
obtain fair fits (although for $\xi<10$ we had to discard the $\NR=32$ data
from the fits). The resulting exponents $B(\xi)$ are shown in
Fig.~\ref{fig:exp_vs_xi}. As it can be checked, the hypothesis $B(\xi)\propto
\log(\xi)$ is tenable, particularly for $\xi>10$.

\section{The \boldmath $\tau(q)$ at $T_\mathrm{c}$}
We have fitted our data for $\tau(q)$ following the same functional forms that
we used at $T=0.9$, namely:
\begin{equation}\label{SI-eq:tau-ansatz}
\tau_1(q)=mq\frac{1+c_1q}{1+c_2 q}\,,\quad \tau_2(q)=mq\frac{1+d_1q \log
  q}{(1+d_2 q)^2}\,.
\end{equation}
Both $\tau_1$ and $\tau_2$ have the same derivative at $q\!=\!0$, namely
$m$. We do not regard $m$ as a fitting parameter. Rather we take it from the
scaling of the \emph{median} of the distribution $P(C_4)$ with $\xi$. As
usual, we first fitted the median as a function of $\overline{C_4}$, see
Fig.~\ref{fig:mediana_vs_C4}, and then translated it to a power law in $\xi$.

\begin{figure}
\centering \includegraphics[width=\linewidth]{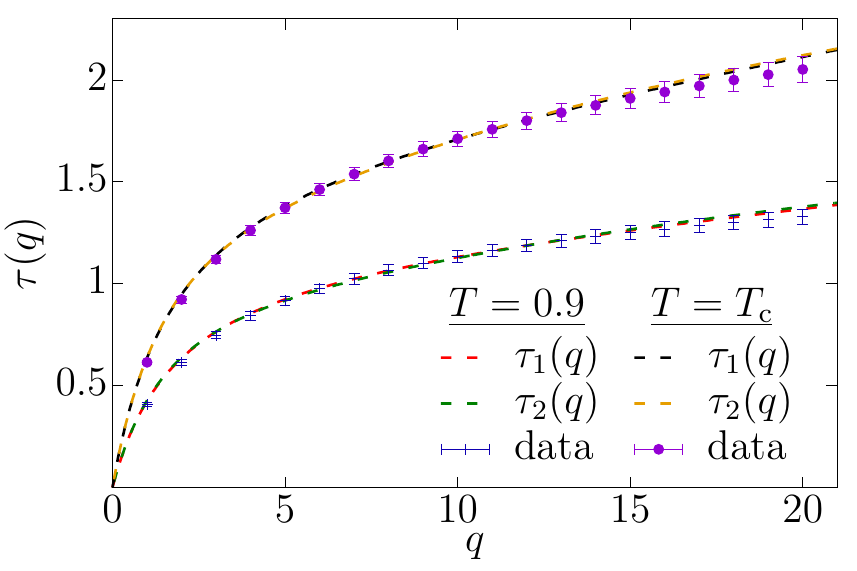}
\caption{The scaling exponent for the $q$-th moment $\tau(q)$, \emph{i.e.},
  $\overline{C_4^q}\sim 1/[\xi(\tw)]^{\tau(q)}$ versus $q$, as computed for
  our simulations of the Ising spin glass at
  $T_\mathrm{c}=1.1019(29)$~\cite{janus:13} and $T=0.9$. The dashed lines are
  fits to the functional forms in~\eqref{SI-eq:tau-ansatz}. Full fit results are
  shown in Table~\ref{tab:fit_summary}.}
\label{fig:tau_q_Tc}
\end{figure}

Fig.~\ref{fig:tau_q_Tc} shows our estimations and fits for $\tau(q)$ for
$T=0.9$ and $T=T_{\mathrm{c}}$. Although the functional behaviors of both
curves are similar, the clear difference reveals the significantly different
nature of the multifractal behaviors at the critical points and in the
spin-glass phase.

\section{Summary of the statistical information fits}

As stated in the main text, and in the above sections, in order to compute and
interpolate $\tau(q)$, and its errors \citep[see][]{yllanes:11}, we need to
make a huge number of fits. For example, to obtain the data shown in
Fig.~\ref{fig:mediana_vs_C4}, we need to extrapolate to $\NR\rightarrow\infty$
the median of $C_4$, see Fig.~\ref{fig:mediana_vs_NR}, 16 times at each
$\xi(\tw)$ in order to estimate errors. This process would be impossible
without some automation. To achieve this, we use a nonlinear least-squares
method using the Levenberg-Marquardt solver~\cite{levenberg:44} routine
available at the GNU Scientific Library. However, in order to calibrate the
process and obtain a fair fits, we perform manually all the fits shown in the
present work. The complete statistical information of the fits is summarized
in Table~\ref{tab:fit_summary}.

\begin{figure}
\centering
\includegraphics[width=\linewidth]{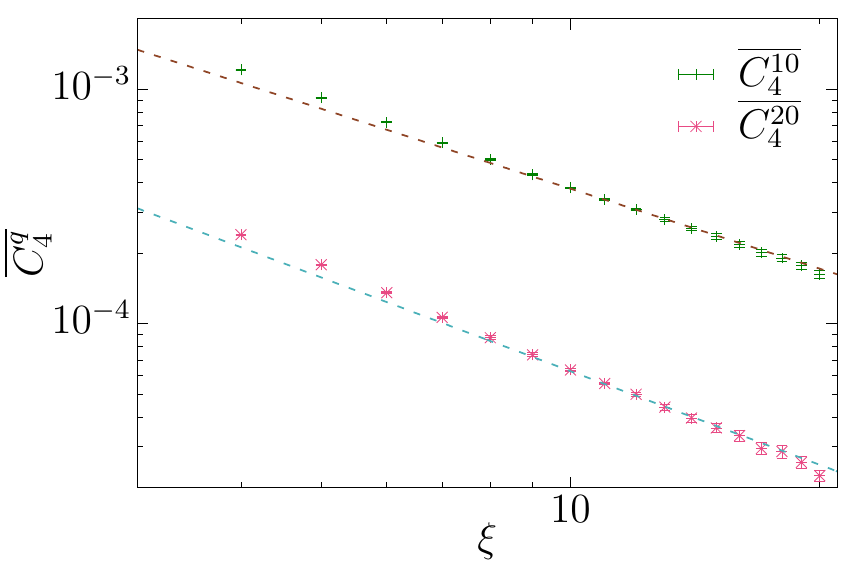}
\caption{The 10th and 20th moments of $C_4$ as functions of the coherence
  length $\xi$ at $T=0.9$.  Dashed lines are fits to $a\xi^{-\tau(q)}$ where
  the only fitting parameter is the proportionality constant $a$ (fit details
  are available in Table~\ref{tab:fit_summary}). The values of $\tau(q=10)$,
  and $\tau(q=20)$ are our final estimates presented in
  Fig.~\ref{fig:tau_q_Tc}. These fits can be considered as a test of our
  analysis, showing that our final estimate of $\tau(q)$ is accurate.}
\label{fig:ajustes_imom10_20_vs_xi}
\end{figure}

Fig.~\ref{fig:ajustes_imom10_20_vs_xi}, which shows the evolution of the 10th
and 20th moments of $C_4$ as function of the coherence length, includes a fit
to a power law $a_q\xi^{-\tau(q)}$ , where the only parameter is the amplitude
constant, $a_q$, and the respective value of $\tau(q)$ is the result of our
analysis. These fits, included in Table~\ref{tab:fit_summary}, are a good
confirmation of our analysis.

\subsection{Interpolations necessary to Fig. 4 in the main text}
Finally, let us recall that Fig.~4 in the main text requires the evaluation of
$C_4^{\text{av}}(r,\tw)$ at non-integer values of its argument. Here, we
shall address \emph{only} the interpolation for our data for the spin glass at
$T=0.9$ and $\tw$ such that $\xi(\tw)=20$.

In order to interpolate our data, obtained for integer $r$, we have performed
a fit that should account for both the short- and long-distance behavior of
the correlations:
\begin{eqnarray}\label{eq:C4-AV-interpolation}
C_4^{\text{av}}(r,\tw)&=&F_\text{sd}(r)\ +\ F_\text{ld}(r)\,,\nonumber\\
F_\text{sd}(r)&=&(b_0+b_1r+b_2r^2)\mathrm{e}^{-(r/2.6)^4}\,,\\
F_\text{ld}(r)&=&\frac{A}{r^\alpha}\mathrm{e}^{-(r/\xi_\text{exp})^\beta}\,.\nonumber
\end{eqnarray}
While the functional form for $F_\text{ld}(r)$ is well known and physically
motivated~\cite{janus:08b,janus:09b,janus:18}, we regard $F_{\text{sd}}(r)$ as
a purely ad-hoc fitting device. The above functional form fits our integer-$r$
data (within errors) for all $r\geq 1$. Of course, the fitting function can be
evaluated whether the argument is integer or not. The fit's figure of merit is
$\chi^2/\mathrm{dof}=9.18/71$ ($\mathrm{dof}$ stands for `degrees of
freedom`), where we have considered only the diagonal part of the covariance
matrix. This partly explains the exceedingly small value of $\chi^2$ that we
obtained in the fit (the departure of the $p$-value from one is $\sim
10^{-18}$). In order to compute $\chi^2$, we considered as well the first
image at $L-r$~\cite{fernandez:19}, \emph{i.e.}, we compared the numerical
data to
$f_{\text{sd}}(r)+f_{\text{sd}}(L-r)+f_{\text{ld}}(r)+f_{\text{ld}}(L-r)$.

We conclude this paragraph with the fit parameters (we report many digits for
the sake of reproducibility). For the short-distance piece we have:
\begin{eqnarray}
  b_0&=&0.0453360389027432\,,\nonumber\\
  b_1&=&-0.0264096080489446\,,\\
  b_2&=&0.00558776268187863\,.\nonumber
\end{eqnarray}
The parameters of the long-term decay are:
\begin{eqnarray}
  \alpha&=&0.45829\,,\quad \beta=1.41217\,,\\
  A&=&0.551495\,,\quad \xi_{\text{exp}}=20.5207\,. \nonumber
\end{eqnarray}

\subsection{Comparisons between the diluted ferromagnet and the spin glass}
\begin{figure}
\centering
\includegraphics[width=\linewidth]{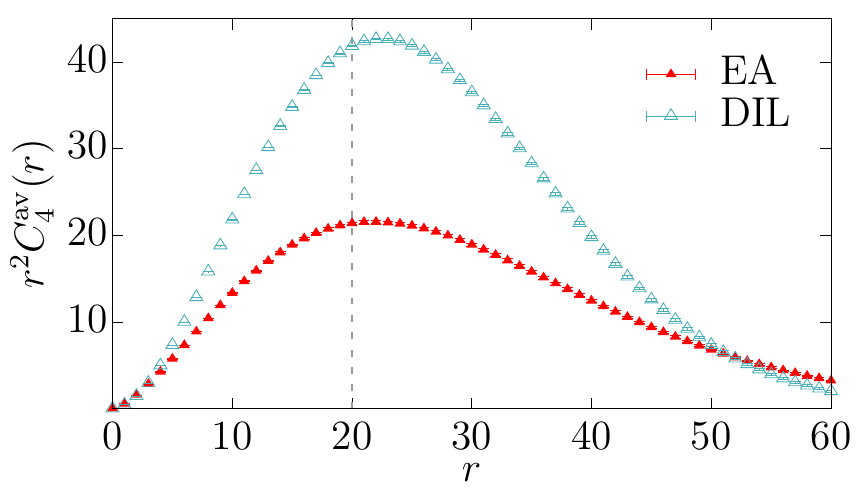}
\caption{Correlation function $C_4^{\text{av}}(r,\tw)$ multiplied by
  $r^2$, as computed for the three-dimensional Ising diluted ferromagnet (DIL)
  and for the Ising spin glass (EA), versus distance $r$. Data were obtained
  in systems of linear size $L=160$ with coherence length $\xi(\tw)=20$
  (dashed vertical line) at temperature $T=0.9$---recall that
  $T_{\mathrm{c}}\approx 1.1$ for the spin glass~\cite{janus:13}. As explained
  in \textbf{Methods} in the main text, the coherence length is computed from
  the integral $I_2=\int_0^\infty r^2C^{\text{av}}_4(r,\tw)\mathrm{d}r$. For both models, $r^2C^{\text{av}}_4(r)$ peaks near $\xi(\tw)$.}
\label{fig:C4_r2_norm_vs_xi}
\end{figure}

\begin{figure}
\centering
\includegraphics[width=\linewidth]{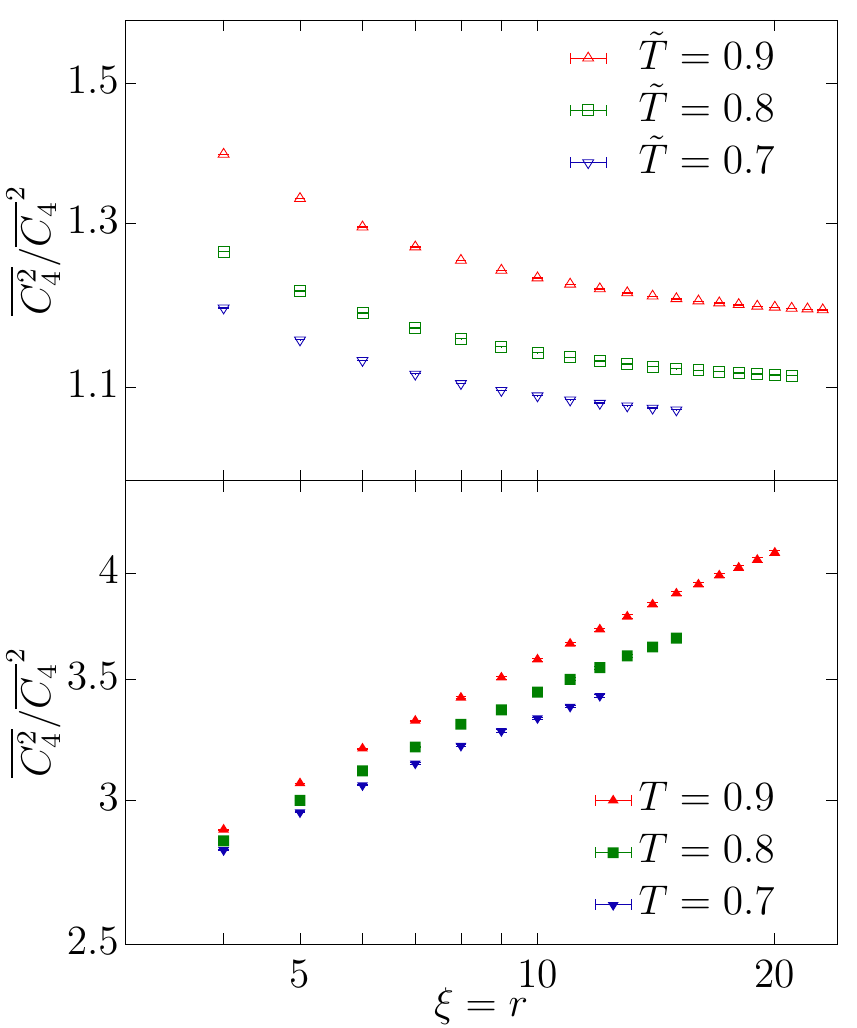}
\caption{Ratio of the second moment of the spin-glass correlation function
  $C_4$ computed at $r=\xi(\tw)$, $\overline{C_4^2}$, to the squared first
  moment, $\overline{C_4}^2$, as a function of the coherence length
  $\xi(\tw)$. Data computed for the Ising diluted ferromagnet (\textbf{top})
  and for Ising spin glass (\textbf{bottom}) at temperatures
  $T,\ \tilde{T}=0.9,0.8$ and $0.7$ (remember that $T^{\mathrm{DIL}}
  =\tilde{T}\,(T_{\mathrm{c}}^{\mathrm{DIL}}/T_{\mathrm{c}})$. Error bars are
  smaller than the point size. Note that, while the diluted ferromagnet tends
  to one for large coherence length, the spin glass follows a power law, which
  indicates that in the scaling limit (\emph{i.e.},
  $\xi(\tw)\rightarrow\infty$) the order of magnitude of $\overline{C_4^2}$ is
  larger than that of $\overline{C_4}^2$. }
\label{fig:C4_norm_vs_xi}
\end{figure}

As explained in the main text, both the diluted ferromagnetic Ising model and
the spin glass display domain-growth dynamics in their low-temperature
phases. In both cases, the size of the domains can be characterized through a
coherence length $\xi(\tw)$, see Fig.~\ref{fig:C4_r2_norm_vs_xi}. However,
while the spin-glass correlation function at distances $r=\xi$ shows strong
statistical fluctuations, see Fig.~\ref{fig:C4_norm_vs_xi}---bottom, this is
not the case for the diluted ferromagnet
Fig.~\ref{fig:C4_norm_vs_xi}---top. Indeed, for the diluted ferromagnet we see
from the figure that $\overline{C_4^2}/\overline{C_4}^2$ quickly reaches a
finite limit as $\xi(\tw)$ grows. Instead, a steady increase of the ratio
$\overline{C_4^2}/\overline{C_4}^2$ is observed for the spin glass.

\renewcommand{\arraystretch}{1.75}
\begin{table*}
  \centering
  \footnotesize
    \begin{tabular}{|c|c|c|r|r|}\hline
Identifier & Functional Form & Fitting Range & $\chi^2/\mathrm{dof}$ &
Parameters\\\hline $[C_4]$, Fig.~\ref{fig:mediana_vs_NR} &
\multirow[c]{2}{*}{$f(x)=a+bx+cx^2$} & \multirow[c]{3}{*}{$\NR \in [32,512]$}
& $0.872768/2$ & $a=0.01492(12),\ b=-0.52(2),\ c=-5.0(6)$
\\ \cline{1-0}\cline{4-5} $[S_{\bsym{x}}S_{\bsym{x}+\bsym{r}}]^2$,
Fig.~\ref{fig:mediana_vs_NR} & & & $0.946227/2$ &
$a=0.01492(12),\ b=0.46(2),\ c=-4.3(6)$ \\\cline{1-2} \cline{4-5} ${\cal
  I}(\NR)$, Fig.~\ref{fig:exp_vs_NR} & $f(x) = ax^{(b+1)}$ & & $3.70448/3$ &
$a=14(2)$, $b=2.02(4)$ \\\hline $B(\xi)$, Fig.~\ref{fig:exp_vs_xi} &
$f(x)=a+b\log x$ & \multirow[c]{3}{*}{$\xi \in [10,20]$} & $5.95738/9$ &
$a=0.78(11)$, $b=0.39(4)$ \\\cline{1-2} \cline{4-5} $\overline{C_4^{q=10}}$,
Fig.~\ref{fig:ajustes_imom10_20_vs_xi} &
\multirow[c]{2}{*}{$f(x)=ax^{-\tau(q)}$} & & $5.6406/10$ & $a=0.00513(2)$
\\\cline{1-0} \cline{4-5} $\overline{C_4^{q=20}}$,
Fig.~\ref{fig:ajustes_imom10_20_vs_xi} & & & $7.64613/10$ & $a=0.001337(11)$
\\\hline Median of $C_4$, Fig.~\ref{fig:mediana_vs_C4} &
\multirow[c]{3}{*}{$f(x)=ax^{b}$} & $\overline{C_4}(\xi\in[4,20])$ &
$1.86484/15$ & $a=1.489(16)$, $b=1.575(4)$\\\cline{1-0}\cline{3-5}
$\text{median}[C_4]/\overline{C_4}$, Fig.~5 & & $\overline{C_4}(\xi\in[4,20])$
& $8.88094/15$ & $a=1.488(7)$, $b=0.575(2)$\\\cline{1-0}\cline{3-5}
$C_4^{\text{av}}(r=\xi; T=0.9)$, Fig. 2 & & $\xi \in [10,20]$ &
$1.99859/9$ & $a=0.182(2)$, $b=-0.411(5)$\\\hline $\tau_1(q)$ for $T=0.9$,
Fig.~5 and Fig.~\ref{fig:tau_q_Tc} &
\multirow[c]{2}{*}{$f(x)=mx\frac{1+ax}{1+bx}$} &
\multirow[c]{4}{*}{$q\in[1,20]$} & $5.01227/18$ & $a=0.0129(18)$,
$b=0.547(12)$ \\\cline{1-0}\cline{4-5} $\tau_1(q)$ for $T=T_{\mathrm{c}}$,
Fig.~\ref{fig:tau_q_Tc} & & & $5.37753/18$ & $a=0.1718(17)$,
$b=0.574(10)$\\\cline{1-2}\cline{4-5} $\tau_2(q)$ for $T=0.9$, Fig.~5 and
Fig.~\ref{fig:tau_q_Tc} & \multirow[c]{2}{*}{$f(x)=mx\frac{1+ax}{(1+bx)^2}$} &
& $9.33556/18$ & $a=0.039(2)$, $b=0.232(5)$ \\\cline{1-0}\cline{4-5}
$\tau_2(q)$ for $T=T_{\mathrm{c}}$, Fig.~\ref{fig:tau_q_Tc} & & & $8.23336/18$
& $a=0.0446(2)$, $b=0.242(4)$ \\\hline
    \end{tabular}
    \caption{Summary of the statistical information of all the fits reported
      in our work. The fits has been done with GNUplot. We have used this data
      in order to calibrate our analysis, which has been automated to estimate
      the errors following the strategy of~\cite{yllanes:11}. }
    \label{tab:fit_summary}
\end{table*}

%

\end{document}